\documentclass[fleqn,10pt]{wlscirep}
\usepackage[utf8]{inputenc}
\usepackage[T1]{fontenc}
\usepackage{epsfig}
\definecolor{mypurple}{rgb}{0.7,0.3,0.8}
\definecolor{midgreen}{rgb}{0.1,0.6,0.3}

\title{Exploration of the effects of epidemics on the regional socio-economics: a modelling approach}

\author[1]{Jan E. Snellman}
\author[2]{Rafael A. Barrio}
\author[1,3]{Kimmo K. Kaski}
\author[4,1,5,*]{Maarit J. Korpi--Lagg}
\affil[1]{Department of Computer Science, Aalto University School of Science, FI-00076 AALTO, Finland}
\affil[2]{Instituto de F{\'i}sica, Universidad Nacional Aut{\'o}noma de  M{\'e}xico, 01000 M{\'e}xico D.F., Mexico}
\affil[3]{The Alan Turing Institute, 96 Euston Rd, Kings Cross, London NW1 2DB, UK}
\affil[4]{Max-Planck-Institut f\"ur Sonnensystemforschung, Justus-von-Liebig-Weg 3, D-37077 G\"ottingen, Germany}
\affil[5]{Nordita, KTH Royal Institute of Technology \& Stockholm University, Hannes Alfv\'ens v\"ag 12, SE-11419, Sweden}

\begin{abstract}

Pandemics, in addition to affecting the health of populations, can have huge impacts on their social and economic behavior. These factors, on the other hand, have the potential to feed back to and influence the disease spreading.  It is important to systematically study these interrelations, to determine which ones have significant effects, and whether the effects are adverse or beneficial. Our recently developed epidemic model with agent-based and geographical elements is used in this study for such a purpose. We perform an extensive parameter space exploration of the socio-economic part of the model, including factors like the attitudes (called values) of the agents towards the disease spreading, health, economic situation, and regulations by government agents. We search for prominent patterns from the resulting simulated data using basic classification tools, namely self-organizing maps and principal component analysis. We seek to isolate the most important value parameters of the population and government agents influencing the disease spreading speed and patterns, and monitor different quantities of the model output, such as infection rates, the propagation speed of the epidemic, economic activity, government regulations, and the compliance of population. Out of these, the ones describing the epidemic spreading were resulting in the most distinctive clustering of the data, and they were selected as the basis of the remaining analysis. We relate the found clusters to three distinct types of disease spreading: wave-like, chaotic, and transitional spreading patterns. The most important value parameter contributing to phase  changes between these phases was found to be the compliance of the population agents towards the government regulations. Our thorough mapping of the model parameters confirms our earlier hypotheses. In compliant populations, the infection rates are significantly lower and the infection spreading is slower, while the population agents' health and economical attitudes show a weaker effect. 
\end{abstract}
\begin{document}

\flushbottom
\maketitle

\thispagestyle{empty}

\section*{Introduction}

Over the past couple of years the Covid-19 pandemic has not only changed our every day habits but also entire way of life. This is because the pandemic apart from being a population wide health issue has also turned out to have a huge impact on our social and economic behaviour. Consequently from the mitigation point of view, there is a need to understand the interrelation e.g. between the non-pharmaceutical restriction measures issued by the government or local authorities and their resulting effects on the infectious disease spreading as well as consequences on the socio-economic activity of people. In this an important element is the degree in which people comply with the government or local authority restrictions like confinement, social distance, closure of schools and commercial activities, etc. 

In order to describe this complex interplay between an epidemic, local authorities, economy and people's behavior, we have recently presented a proof of concept model combining socio--economic simulations with epidemic modeling, called the BTH-SEIRS model \cite{SBKK2022}. The epidemic part of this hybrid model is based on a SEIRS system \cite{KK1991a,KK1991b,KK1991c} and it takes into account separately the  properties of the disease, namely its virulence and lethally, with stochastic mechanisms for the  geographical spread \cite{BVGM2013,BKHAG2020,BGBB2020,BVGNBB2021}, while the social part was based on an agent--based implementation of the better--than hypothesis (BTH, introduced in \cite{SGGBK2017,SGKBK2018}) which takes into account the decision power of individuals through an appropriate decision function.  The main motivation of the model was to create a rudimentary bridge between pandemic modeling and simulating the behaviour of human societies with a simple psychological model. While there have been  agent--based models concentrating on the economic effects of epidemics \cite{BDHHK2020,SBLAGS2020,KSJJ2021}, these have tended not to take into account the reactions that human beings naturally have towards spreading diseases. The raison d'etre of BTH--SEIRS model is to study the socio--economic and public health consequences of an epidemic in a simple simulated society that reacts to it and can take measures to mitigate its spread.

The social part of the model has a total of six different parameters associated with the values of populations and governments, which have various effects on the pandemic spread \cite{SBKK2022}: the populations value their own  efforts to mitigate the epidemic, their health status and their compliance with the restrictions put in place by the governments, while the governments value the health and economic situations in their districts. Of these values the compliance of the populations turned out to have most impact on the spreading of the epidemic, with low or negative compliance leading to the  epidemic spreading very fast in wavelike patterns, while high compliance resulted in the epidemic spreading slowly and chaotically. However, as these demonstrated effects were based on relatively few simulations, they raised more questions, such as to what regions of the parameter space are these behavioural types limited to, are there transitional regions between these types and what are the possible transitional behaviours like, and whether completely new kinds of behavioural types can be found in the six dimensional parameter space. In this study we set out to answer these questions.

In order to reduce the dimensionality of the parameter space for this study, we re-wrote the equations in a different form, which enabled us to reduce the dimensionality of the parameter space from six to four. Even so, we had to perform a total of $10^5$ simulations to map an adequate parameter space. Obviously, these simulations produced a vast amount of data, and analysing them was not possible manually, but the analysis pipeline had to be automated. As we were looking for the most prominent patterns in the data, for analysis tools we selected the principal component analysis (PCA \cite{P1901,H1933,H1936}), self-organising maps (SOM \cite{2004KSOM}) and silhouette numbers \cite{R1987}.We aim at determining whether the feature vectors formed from the simulation data exhibit clusters, and whether they are significant. 

The rest of this paper is organised as follows: In the Methods--section we present the model, the parameter space and how the simulations are analysed. In the Results--section we concentrate on the main findings, while we relegate some of the more mundane results to an appendix. the significance of the results are discussed in the final section.

\section*{Methods}
\label{modsec}

In \cite{SBKK2022} we presented an epidemic model with agent-based and geographical elements. The modeled epidemic spreads in a geographical area divided into governmental districts, which in turn are subdivided into much smaller geographical cells. The authorities of the governmental districts and the populations of the cells are treated as agents that influence the spreading of the epidemic with their economic policies: restricting economic activity reduces the probability of the epidemic spreading from one cell to its neighbours. Within the cells the evolution of the epidemic follows the SEIRS-dynamics. The behavioural model of the agents is based on BTH, which allows for the effects of the values of the agents to be included. The authority agents place value in the restrictions they put in place to mitigate the epidemic and the overall health and economic activity in their districts, while the population agents place value in their own effort to mitigate the epidemic, their own health and their compliance with the authorities' restrictions. In this study we make massive amounts of simulations with this model using our simple original setup from \cite{SBKK2022}: a total of nine districts comprising of $17 \times 17$ geographical cells in $3 \times 3$ grid pattern.

The behaviour of the agents are based on the following utility equations
\begin{eqnarray}
u_i &=& w^x_i (x_i + \sum_{j \in e_i} (x_i - x_j)) 
 + w^y_i (y_i + \sum_{j \in e_i} (y_i - y_j)) 
 + w^c_i (c_i + \sum_{j \in e_i} (c_i - c_j)), \\
U_i &=& W^X_i (X_i + \sum_{j \in E_i} (X_i - X_j)) 
 + W^Y_i (Y_i + \sum_{j \in E_i} (Y_i - Y_j)) 
 + W^Z_i (Z'_i + \sum_{j \in E_i} (Z_i - Z_j)),
\label{govpopUeq}
\end{eqnarray}
where the lower case letters refer to the population agent measures and capital letters to authority agent measures. Thus, 
\begin{itemize}

    \item $u_j$ and $U_k$ are the BTH utilities of the population agent $j$ and authority agent $k$, respectively.
    
    \item $x_j$ and $X_k$ are the reduction of economic activity by the population agent $j$ and restrictions put in place by authority agent $k$, respectively. The probability of the epidemic spreading from a cell to its neighbour is $\nu^t = v_0 - x_i v_{max}$, where $v_0 = 0.5$ is the base spreading probability and $v_{max} = 0.49$ is its maximum reduction achievable with non--pharmaceutical interventions (NPIs). The values for $v_0$ and $v_{max}$ have been chosen arbitrarily with demonstration purposes in mind.
    
    \item $y_j$ and $Y_k$ are the infection rates in the cell of population agent $j$ and the overall infection rated in the district governed by authority agent $k$, respectively.
    
    \item $c_j = x_j - X^r_j$ and $Z_k = \sum_{j \in D_k } x_j$ are the compliance of the population agent $j$ and the overall reduction of economic activity in the district $D_k$ governed by authority agent $k$, respectively. $Z_k$ is linked to the gross domestic product $Z'_k$ by $Z'_k = 1 - Z_k$.
    
    \item $w^x_i$, $w^y_i$ and $w^c_i$ are the values that the population agent $i$ holds for their own effort to mitigate the epidemic, their own health and their compliance with the authorities' restrictions, respectively.
    
    \item $W^X_i$, $W^Y_i$ and $W^c_i$ are the values that the authority agent $i$ holds for the restrictions they put in place to mitigate the epidemic and the overall health and economic activity in their districts, respectively.
    
\end{itemize}
The simulations of this model are run in discrete time steps, and the agents update their behaviour, i.e. the economic efforts of the population agents $x_i$ and the restrictions $X_i$ of the authority agents to mitigate the epidemic, according to the evolving epidemic. With some assumptions, principally that the agents will make the minimum effort to mitigate the epidemic ($u_i=U_i=0$), we derived the following expressions for $x_i$ and $X_i$ for each time step:
\begin{eqnarray}
x_i &=& \frac{1}{|e_i| + 1} ( \frac{w^x_i}{w^x_i + w^c_i} \sum_{j \in e_i} x_j  
 - \frac{w^y_i}{w^x_i + w^c_i} (y_i + \sum_{j \in e_i} (y_i - y_j))) 
 + \frac{w^c_i}{w^x_i + w^c_i} (|e_i| X^r_i - \sum_{j \in e_i} c_j), \\
X_i &=& \frac{1}{|E_i| + 1} (\sum_{j \in E_i} X_j - \frac{W^Y_i}{W^X_i} (Y_i + \sum_{j \in E_i} (Y_i - Y_j)) 
 - \frac{W^Z_i}{W^X_i} (Z_i + \sum_{j \in E_i} (Z_i - Z_j))).
\label{dmeqgovmkI}
\end{eqnarray}
Since the minimum effort assumption implies  $w^c_i < 0$ and $W^X_i < 0$, the above equations can be written in the form
\begin{eqnarray}
x_i &=& \frac{1}{|e_i| + 1} ( \frac{1}{\Tilde{w^c_i} - 1} \sum_{j \in e_i} x_j  
 - \frac{\Tilde{w^y_i}}{\Tilde{w^c_i} - 1} (y_i + \sum_{j \in e_i} (y_i - y_j))) 
 + \frac{\Tilde{w^c_i}}{\Tilde{w^c_i} - 1} (|e_i| X^r_i - \sum_{j \in e_i} c_j), \label{dmeqgovmk} \\
X_i &=& \frac{1}{|E_i| + 1} (\sum_{j \in E_i} X_j + \Tilde{W^Y_i} (Y_i + \sum_{j \in E_i} (Y_i - Y_j)) 
 + \Tilde{W^Z_i} (Z_i + \sum_{j \in E_i} (Z_i - Z_j))),
\label{dmeqgovmkII}
\end{eqnarray}
where 
\begin{equation}
\Tilde{w^y_i} = \frac{w^y_i}{\lvert w^x_i \lvert},
\Tilde{w^c_i} = \frac{w^c_i}{\lvert w^x_i \lvert},
\Tilde{W^Y_i} = \frac{W^Y_i}{\lvert W^X_i \lvert},
\Tilde{W^Z_i} = \frac{W^Z_i}{\lvert W^X_i \lvert}, 
\label{tildes}
\end{equation}
and $w^c_i \neq w^x_i$ (or $\Tilde{w^c_i} \neq 1$) always to prevent division by zero. 

The formulations in Eqs. \ref{dmeqgovmkII} and \ref{tildes} allow us reduce the dimensionality of this parameter survey from six to four, since we need only consider the tilde parameters. Thus we conduct the survey with the following grid:
\begin{itemize}

\item $\Tilde{w^y_i} = -0.25, -0.5, -1, -2, -4,-3, -5, -10, -25, -50, -15, -20, -30, -40, -45$ 
        
\item $\Tilde{W^Y_i} = -0.25, -0.5, -1, -2, -4,-3, -5, -10, -25, -50, -15, -20, -30, -40, -45$ 
        
\item $\Tilde{w^c_i} = -0.25, -0.5, -1, -2, -4, 0.25, 0.5, 1, 2, 4, -0.75, -3, -5, -7.5, -10, 0.75, 3, 5, 7.5, 10$
        
\item $\Tilde{W^Z_i} = -0.25, -0.5, -1, -2, -4, 0.25, 0.5, 1, 2, 4, -0.75, -3, -5, -7.5, -10, 0.75, 3, 5, 7.5, 10$
        
\end{itemize}
We have chosen $\Tilde{W^Y_i}$ and  $\Tilde{w^y_i}$ to be always negative because of the assumption that the population and authority agents will make minimum efforts to mitigate the epidemic, ruling out herd immunity approach to dealing with the epidemic. We have chosen $\Tilde{W^Z_i}$ to have both positive and negative values, even though it means that authority agents in the former case consider a declining economy to be an asset. This choice was made because there are no theoretical reasons to exclude positive values of $\Tilde{W^Z_i}$, although there may be no real world authorities taking this view. 

For each of these gridpoints we made two simulations in which the value parameters were randomized within radius of $0.1$ and the mean value indicated by the gridpoint, for a total of $10^5$ simulations. From these simulations we saved  the information on the number of districts, size of the districts, value parameters of the authority and population agents, epidemic arrival time map $A_{ij}$ and slowness $\beta$, times series of gross domestic products, government regulations, population compliance and infection rates by district. Of these the epidemic arrival time map and slowness are measures of the speed of the spreading epidemic in the simulation, which we have introduced to monitor the spread of the epidemic with more detail in our simulations. These work as follows: The time step at which the epidemic spreads to geographical cell located at $(i,j)$ is recorded to the arrival time map element $A_{ij}$. If the epidemic does not spread to a cell during the time frame of the simulation, the time will be marked down as the last time step of the simulation. The elements of the matrix $A_{ij}$ are then summed together to get an overall measure for the spreading rate of the epidemic,
\begin{equation}
A_{T} = \sum_{i,j} A_{ij}.
\end{equation}
$A_T$ attains its maximum value $A^{max}_T$ in the situation where the epidemic is not spreading at all from its initial beginning cell, in which case 
\begin{equation}
A^{max}_T = T_s \times (N_c - 1) + 1,
\end{equation}
where $T_s$ is the simulation length and  $N_c$ the number of simulated cells. Comparing $A_T$ to $A^{max}_T$ yields a fractional measure that describes the slowness of the epidemic spread 
\begin{equation}
\beta = \frac{A_T}{A^{max}_T},
\end{equation}
which we will call the $\beta$ measure. The lower $\beta$ is, the faster the propagation of the epidemic, with a theoretical minimum value of
\begin{equation}
min(\beta) = \frac{N_c}{A^{max}_T},
\end{equation}
which occurs in the case where the epidemic spreads instantly to all the geographical cells in a simulation, while $\beta = 1$ indicates that the epidemic does not spread at all, as stated above. The $\beta$ measure can be defined for each district separately by taking into account the arrival time of the epidemic into each district.

With the information from the simulations we then endeavour to find different behavioural types in the simulations with self organising maps using the Minisom python package \cite{vettigliminisom}. We also made use of the Scikit Learn python package \cite{scikit-learn} to perform a principal component analysis for the minisom feature vectors for visualisation purposes. We experimented with a variety of feature vectors and found that the ones based on infection rates within districts and the district $\beta$ measures exhibited the most easily recognizable clustering behaviours. In addition to feature vectors the self organising maps are also characterised by the number of neurons used, which determine the maximum number of clusters the method is allowed to find. To evaluate the quality of the clustering found by Minisom with different neuron numbers, we used the standard silhouette numbers.

\section*{Results}
\label{resec}

In this study our main objective is to search for different social, epidemiological or economic behavioural patterns in the value parameter space of the BTH-SEIRS model using computational methods, and determine whether these behavioural types are contained in specific regions of the parameter space, in which case they can be thought as equivalent to phases in physical systems. This approach is motivated by the results of our initial study \cite{SBKK2022}, in which we discovered two very different spreading patterns: a fast, wavelike spreading pattern and a slow, chaotic spreading pattern. Compliance value parameter in particular seemed to have most effect on Which of these two spreading patterns would be present in a simulation. Because the contrast between these two patterns is rather easy to spot, even with naked eyes, we have especially put effort in trying to isolate these behaviours in the parameter space, although not to the complete exclusion of seeking other possible behavioural types. The quantities tracked by our model are the average economic activity, government regulations, dynamic compliance, infection rates in the nine modeled districts, and the $\beta$ measures defined in the previous section, and we construct feature vectors for classification purposes for all of these. However, the last two of these measures are just different ways of tracking the evolution of the epidemic. The point of using two different measures for the same thing is, of course, to see if they agree, which would support the notion that the wavelike and chaotic spreading patterns form their own phases in the value parameter space.

Our methodology can be detailed as follows: For the first four measures listed above we construct the feature vectors by averaging the simulation data over time and then arranging the acquired averages by the district number, so the feature vectors take the form (average measure in district $1$, average measure in district $2$, ..., average measure in district $9$). For the $\beta$ measure we take into account both the last $\beta$ time stamp in each district and the individual district $\beta$ measures and arrange this data into the following feature vector: (maximum recorded $\beta$ time step in district $1$, ..., maximum recorded $\beta$ time step in district $9$, $\beta$ measure in district $1$, ..., $\beta$ measure in district $9$). Next, we perform a dimensional reduction on these vectors using PCA to visualise them in three dimensions, to better make sense of the computational classifications of this data for which purpose we use SOMs. The SOMs require a two dimensional neuron matrix as input, to which the data is mapped. We use $1 \times n$ row matrices as the input, with neuron counts $n$ going from $2$ to $20$. After evaluating the quality of these classifications with silhouette numbers, we then proceed to study closer the physical meanings of the best performing ones. 

In the following subsections we present the results of this procedure in the order given above. It should be noted that we are only interested in the large scale behaviours of the system, and so we use relatively simple methods as the PCA and SOMs to make our classifications, as opposed to more involved methods such as UMAP or t-SNE. Since the classification results are not interesting in themselves, we assign them to an appendix, but in short, they can be characterised as follows: Infection rate and $\beta$ classifications form easily identifiable clusters, while the regulation classification has two clusters touching each other. The economic and compliance classifications consist of bundles of filamentary structures that are not easily clustered. Next we take a closer look on the infection rate, regulation and $\beta$ classifications, while ignoring the messy economic and compliance classifications. As stated above, our goal is to find easily identifiable phases, not to shift through all the possible fine structures present in the data we generated as part of this parameter survey. 

\subsection*{The classification by infection rates and $\beta$ measures}

\begin{figure}[t]
\centering
\epsfxsize = 0.45\columnwidth \epsffile{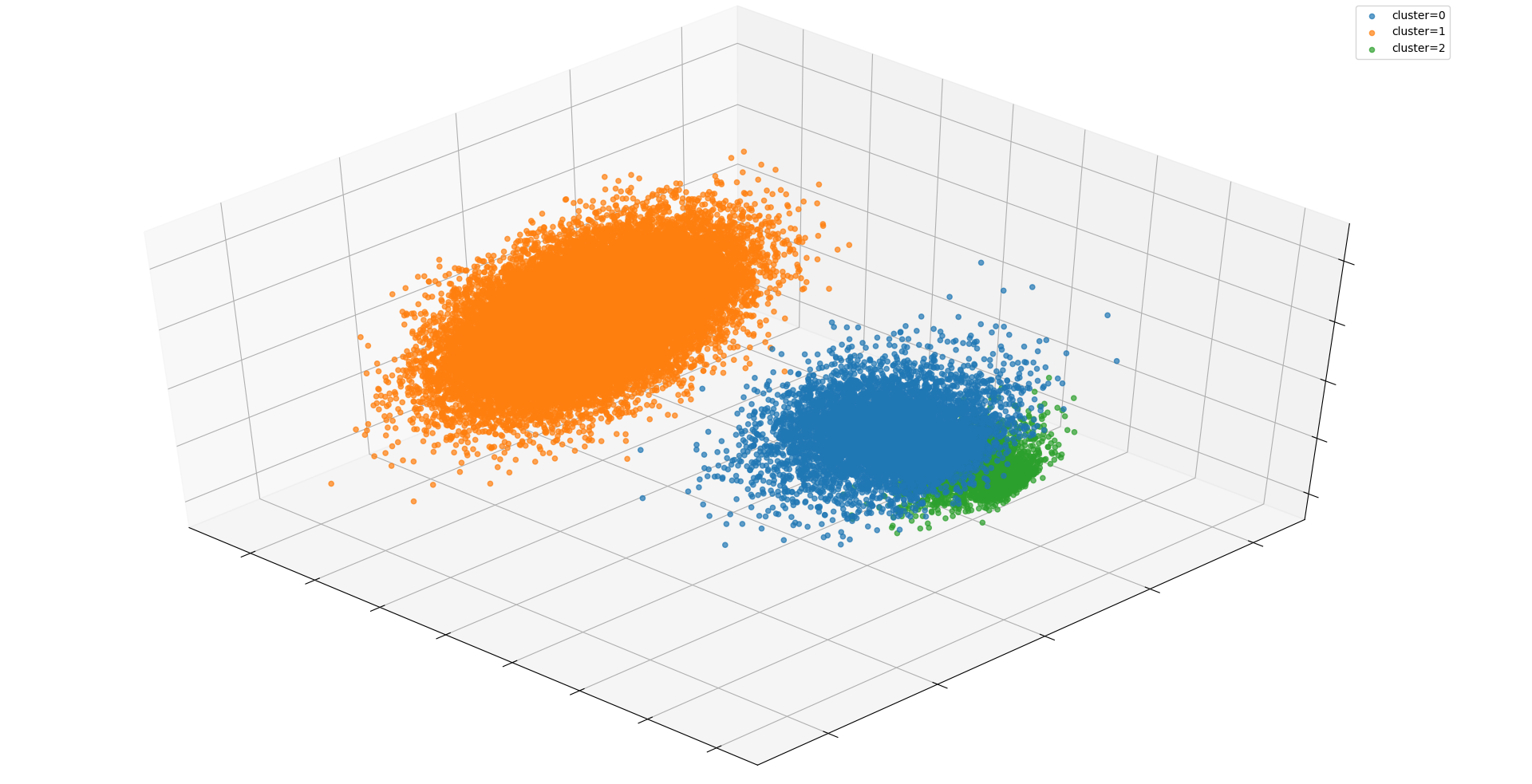}
\epsfxsize = 0.45\columnwidth \epsffile{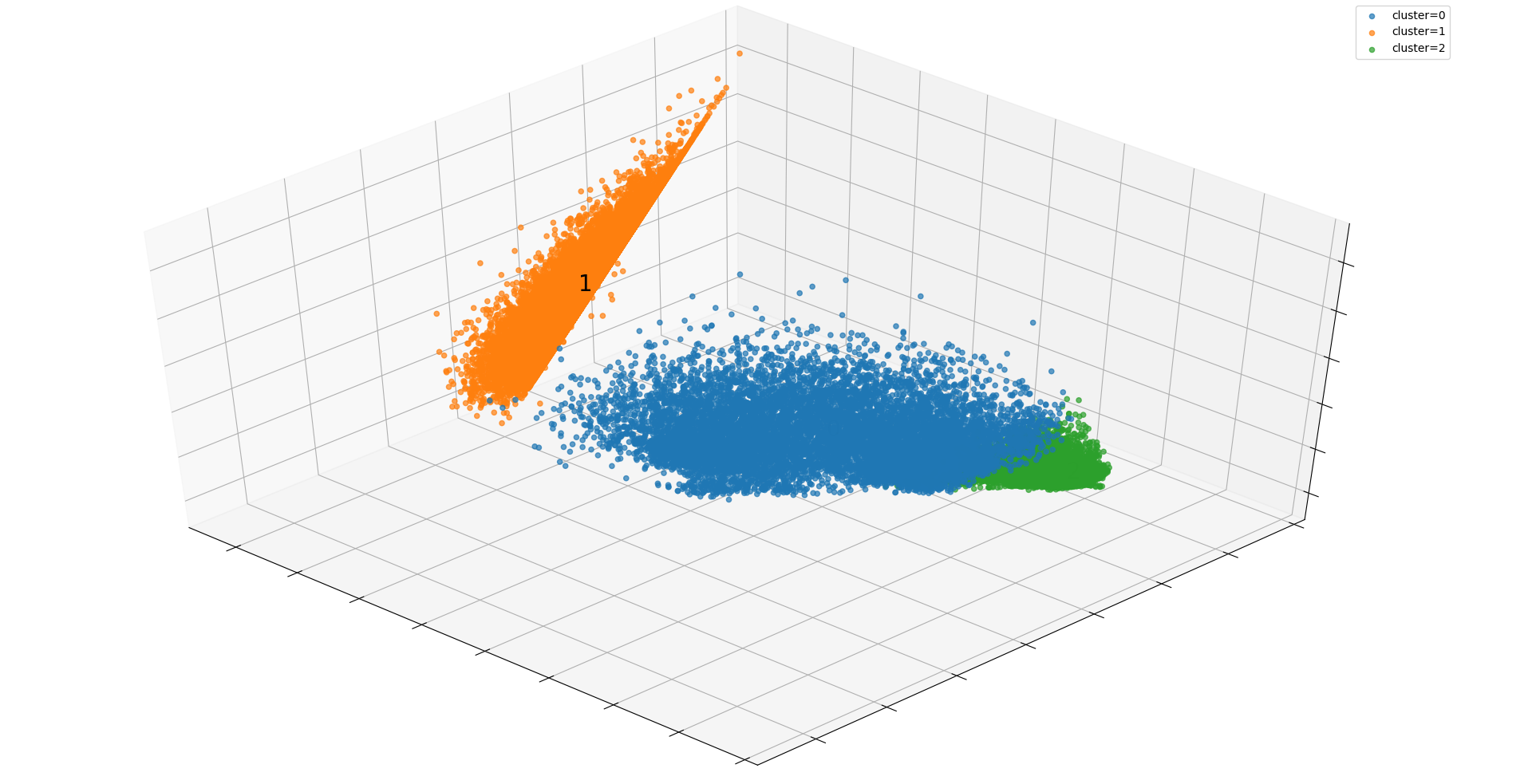}
\epsfxsize = 0.45\columnwidth \epsfysize = 0.15\textheight \epsffile{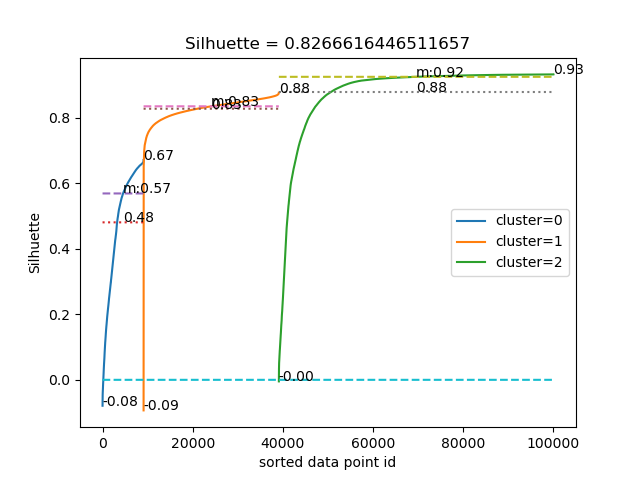}
\epsfxsize = 0.45\columnwidth \epsfysize = 0.15\textheight \epsffile{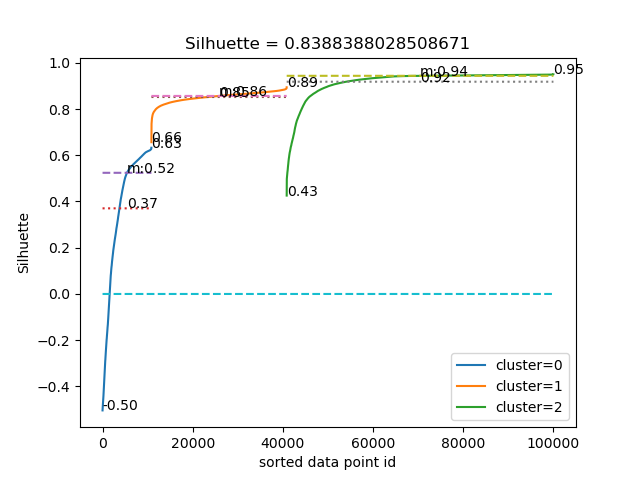}
\epsfxsize = 0.45\columnwidth \epsffile{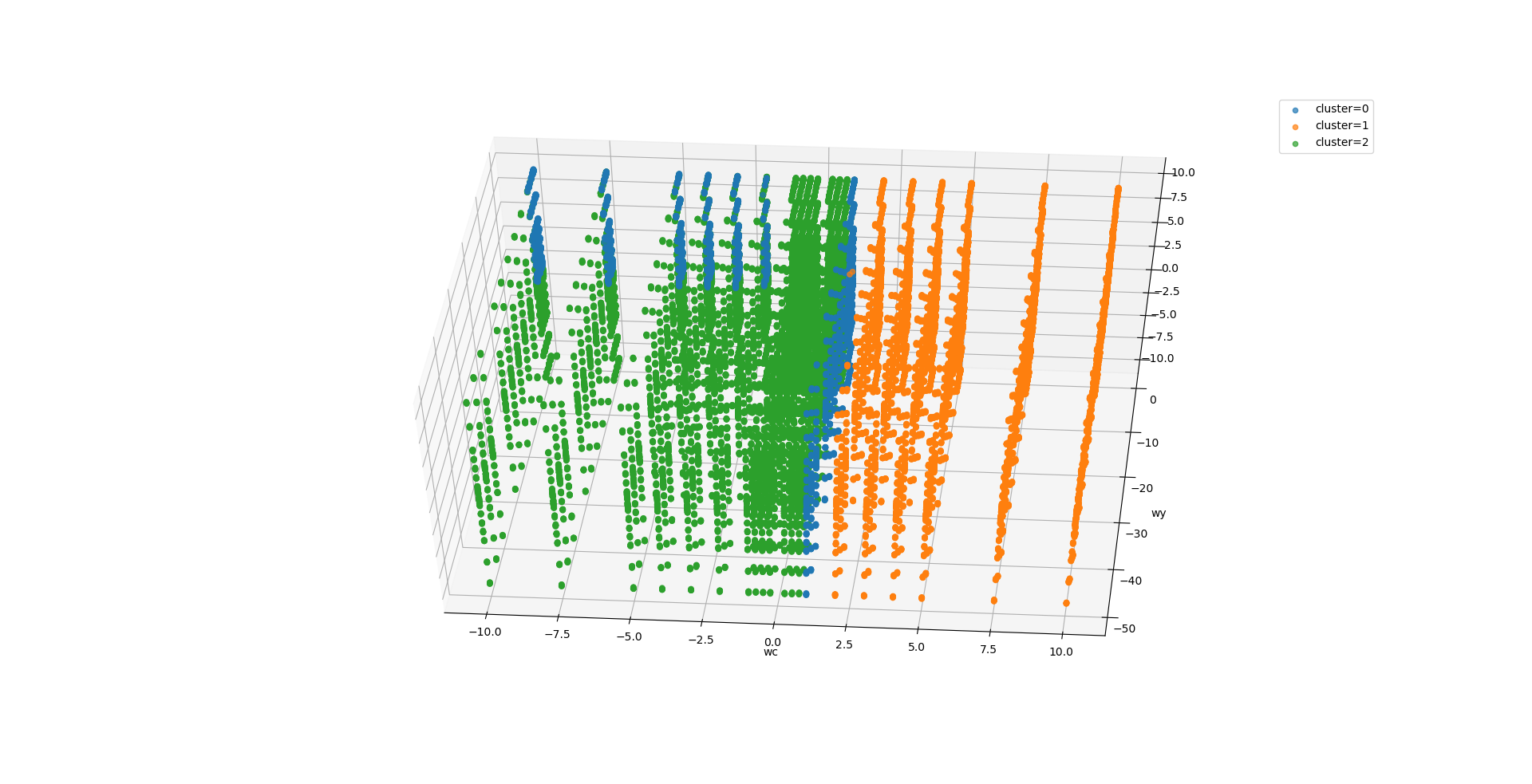}
\epsfxsize = 0.45\columnwidth \epsffile{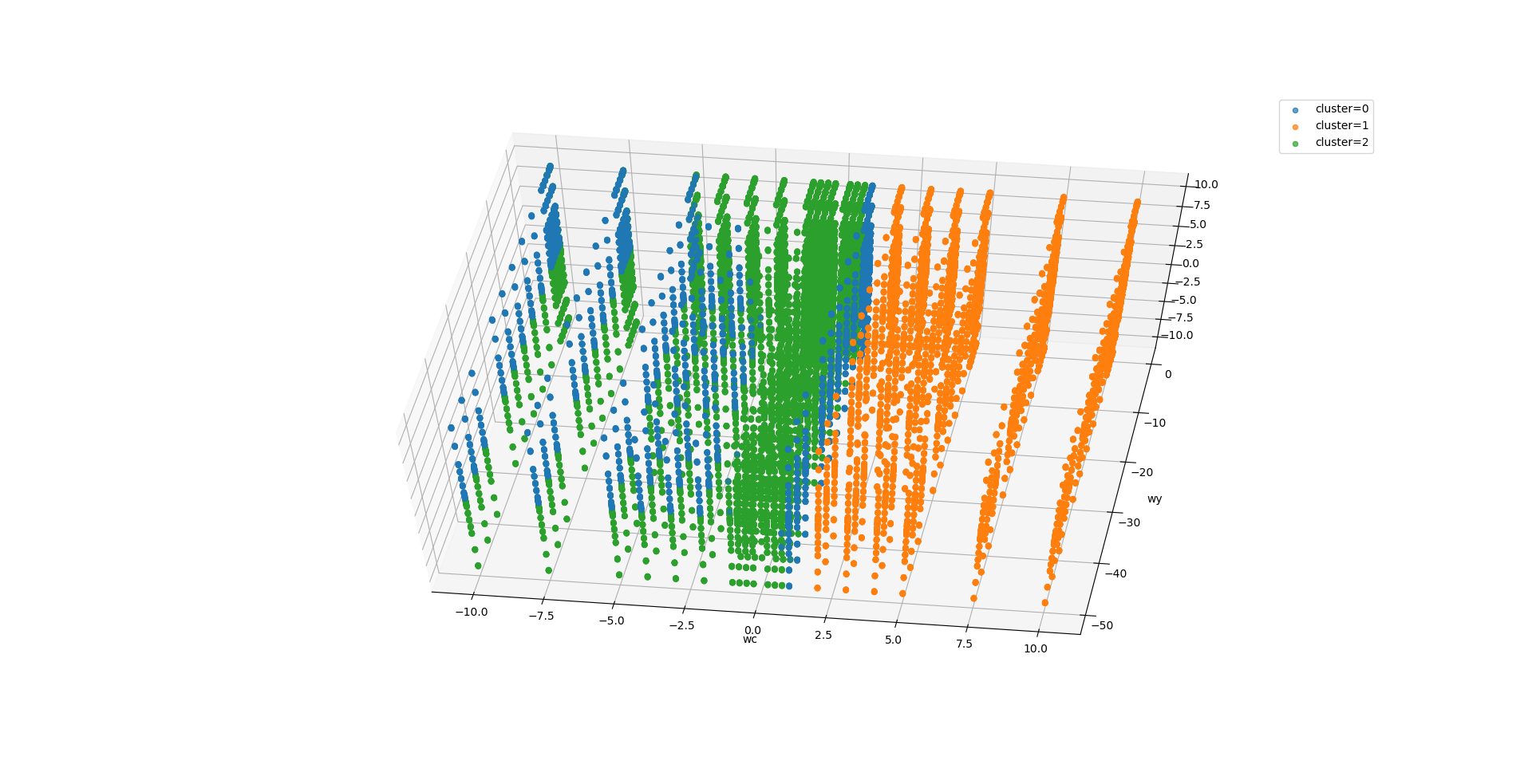}
\caption{The SOM classifications based on average infection rates (top left panels) and the $\beta$ measures (top right panels) with three SOM neurons. The top panels depict the distribution of the SOM clusters as arranged by PCA, the middle panels show the silhouette profiles of the classifications. The bottom panels show the distribution of the clusters in the value parameter space for the infection rate classification: the bottom left panel has $\Tilde{W}^Y$ on the $y$--axis, while the bottom right panel has $\Tilde{w}^y$ on the $y$--axis.
}
\label{Sompcafig1}
\end{figure}

As shown in the appendix, the infection rate and $\beta$ classifications using SOMs produce consistently highest quality results as judged by the silhouette numbers, when compared with other classification schemes we used. In this section we study, why this is the case. It turns out that the easiest way to see this is by taking a closer look on the classifications using three SOM neurons. The top panels of Fig. \ref{Sompcafig1} illustrate this case in PCA space, while the middle panels show the  silhouette profiles. The profiles are constructed by arranging the simulations in the SOM assigned clusters to ascending order according to their silhouette numbers, and then plotting their silhouette number clusterwise. In the profile we have marked the maximum, minimum, average ($a$) and median ($m$)  silhouette numbers of all the clusters, and the width of the lines associated with the clusters indicated the size of the clusters. The positioning of these clusters are shown in the value parameter space in the bottom panels of Fig. \ref{Sompcafig1}, in which two different visualisations are shown due to the inability of showing a four dimensional parameter space at once in a three dimensional space. The difference between visualisations is that the y--axis depicts either $\Tilde{W}^Y$ (top panels) or $\Tilde{w}^y$ (bottom panels), while the x -- and z -- axes are the same for both and stand for $\Tilde{w}^c$ and $\Tilde{W}^Z$, respectively. The value parameter not shown in the figure, either $\Tilde{W}^Y$ or $\Tilde{w}^y$, is collapsed to the three dimensional figure, so the all the simulations are shown in both of the visualisations. Since the clusters are very similar in shape in both the infection rate and $\beta$ classifications, only the former is shown.

The top left panel of Fig. \ref{Sompcafig1} tell us that in the case of infection rate feature vector SOM has, working with three neurons, classified the less dense of the two identifiable PCA clusters as as one cluster (named cluster $1$ in the figure) and divided the denser cluster in to two: dense tip on the right hand side of the cluster, and the rest (named cluster $2$ and cluster $0$, respectively). Similarly, in the case of $\beta$ classification the cluster looking like a sharp wing has been designated a cluster of its own by SOM, while the looser winglike cluster has been divided into two. In the value parameter space, as shown in the bottom panels of Fig. \ref{Sompcafig1}, we see that these clusters mostly occupy the spaces defined by their position on the $\Tilde{w}^c$ axis: Cluster $1$ consists of the simulations on the left side of the $\Tilde{w}^c = 1$ plane, while cluster $2$ takes most of the right hand side of the the $\Tilde{w}^c = 1$ plane, with the exception of a set of simulations with $\Tilde{W}^Z > 0$ and $\Tilde{w}^c \leq -2$, which form a part of cluster $0$ instead. The $\Tilde{w}^c = 1$ plane makes up the rest of cluster $0$.  Henceforth we will call these separate subsections of cluster $0v$ and $0s$, respectively. In terms of the  health value parameters the cluster $0s$ is limited to small values of $\Tilde{W}^Y$  and for the most part large values of $\Tilde{w}^y$, although with large enough values of $\Tilde{w}^c$ the $0s$ subcluster reaches also to the smallest values of $\Tilde{w}^y$.

The silhouette profile of the infection rate classification shows that judging by the mean and median silhouette numbers, cluster $2$ is probably the best defined of these clusters, followed by cluster $1$ and then cluster $0$, while cluster $1$ does have marginally lower minimum silhouette number than cluster $0$. The results for the $\beta$ measure classification are almost identical to the infection rate classification in the value parameter space, while the silhouette numbers indicate that the clusters equivalent to clusters $1$ and $2$ are generally much better defined in the $\beta$ classification, with the third cluster being in turn much worse defined.

\begin{figure}[t]
\centering
\epsfxsize = 0.45\columnwidth \epsffile{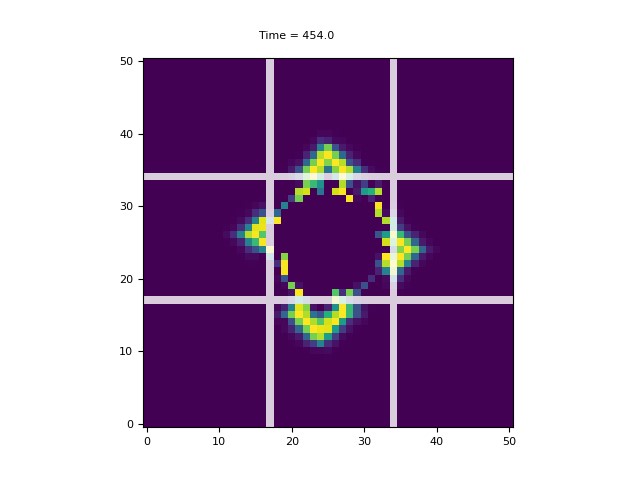}
\epsfxsize = 0.45\columnwidth \epsffile{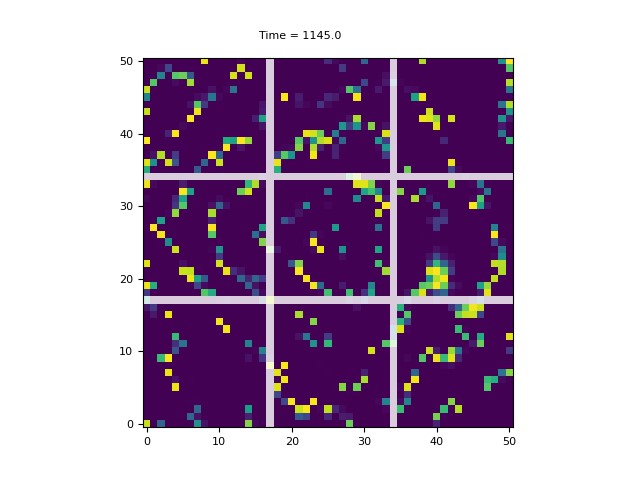}
\epsfxsize = 0.45\columnwidth \epsffile{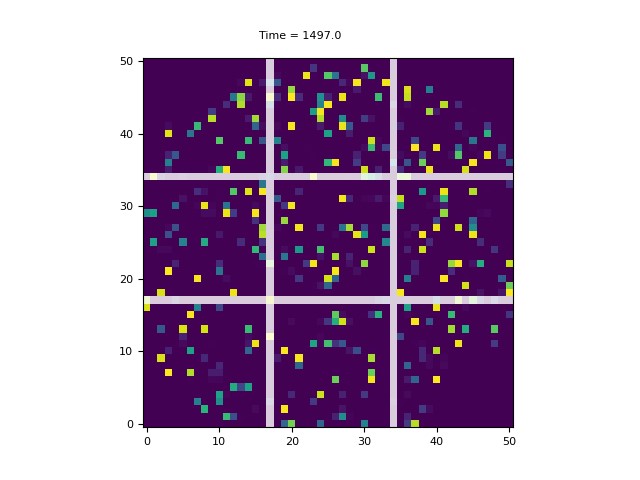}
\epsfxsize = 0.45\columnwidth \epsffile{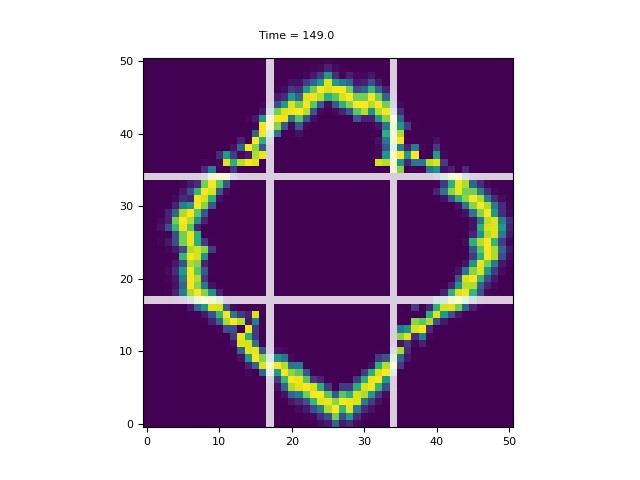}
\caption{Representative visualisations of the simulations near the centroids of the clusters. Two samples for the transitional cluster $0$ are provided: one in between clusters $1$ and $2$ with $w^c = 1$, and the second in in the centroid of those points found in the upper left hand side of Fig. \ref{Sompcafig1} A.}
\label{vissims}
\end{figure}

Having established these clusters with the use of SOM, we have to now determine their physical meanings. The results of \cite{SBKK2022} identified two main behavioural types in the BTH-SEIRS model: the regular and fast spreading wave fronts with negative $\Tilde{w}^c$ and slow and chaotic spreading patterns with positive $w^c$, which clearly correspond to clusters $2$ and $1$ of the infection rate classification, respectively. Just as obviously, cluster $0$, or at least its $\Tilde{w}^c = 1$ component $0v$, represents some form of transitional state between these two behavioural types. Two things should be noted at this point about the effects of increasing or decreasing the number of SOM neurons used to make the infection rate classification, which also apply to the $\beta$ measure classification:

\begin{enumerate}

\item If only two neurons were used to make the classification, clusters $0$ and $2$ (or their equivalents in the $\beta$ classification) will merge into one, while cluster $1$ would remain effectively unchanged, which is the expected result judging from Fig. \ref{pcafig1}.

\item If more than three neurons are used to make the classification, clusters $1$ and $2$ will remain unchanged for the most part, while cluster $0$ will be divided into as many new clusters as the neuron count allows the SOM to find.

\end{enumerate}

These facts indicate that the transitional state from the wave spreading pattern to the chaotic pattern represented by cluster $0$ is very complex, with great deal of local variation according to the value parameters used in each simulation. It is noteworthy that the density of the three clusters seems to decrease in the PCA visualisation as their regularity increases, which is perhaps to be expected as PCA organizes the simulations according to their contribution to the variance between the simulations. At least intuitively one would expect more variability in the chaotic spreading pattern than the wave like pattern.

\begin{figure}[t]
\centering
\epsfxsize = 0.85\columnwidth \epsffile{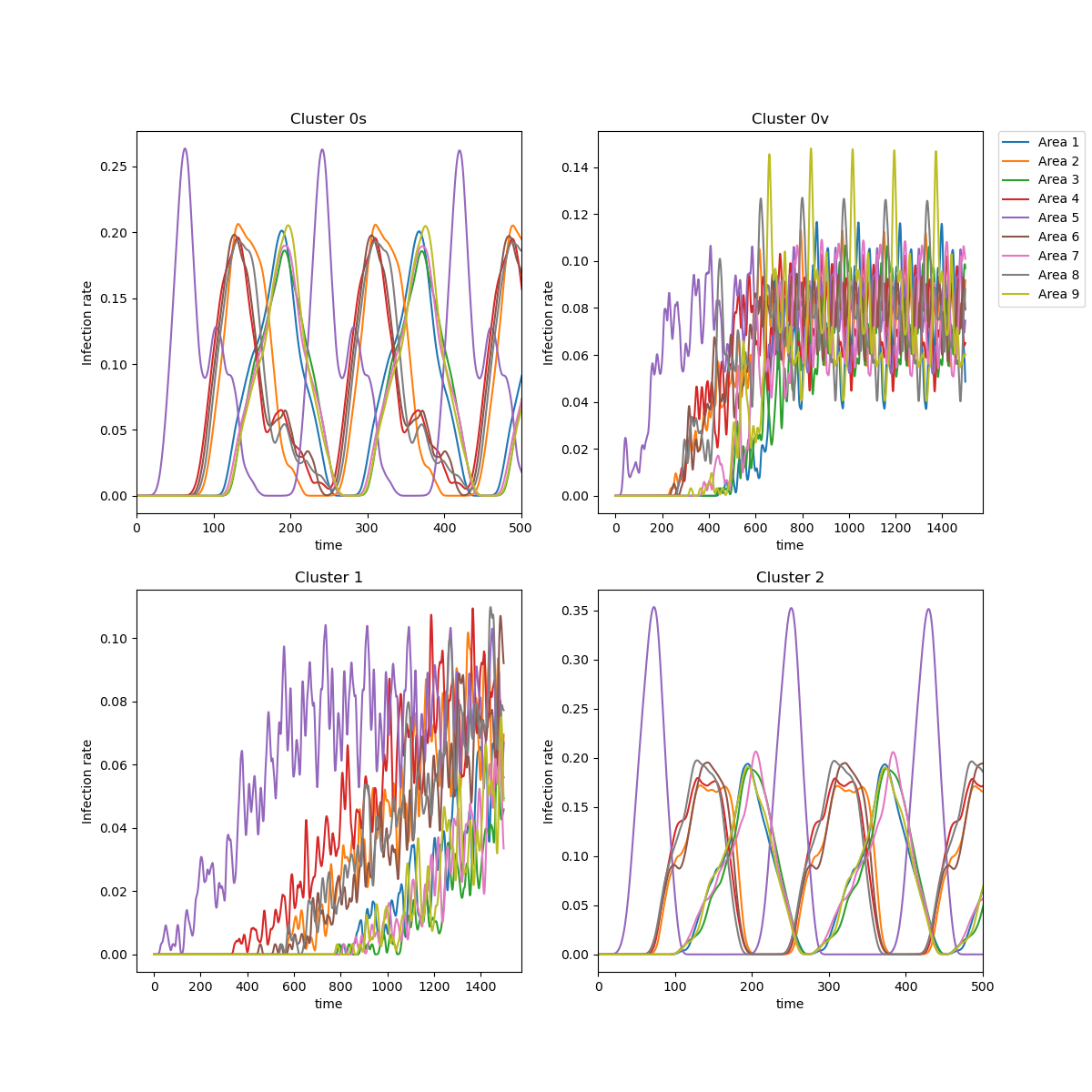}
\caption{Infection rates in different centroidal simulations of the clusters.}
\label{infrates}
\end{figure}

In order to demonstrate the behavioural types of the clusters found by SOM, we performed simulations in the vicinities of the centroids of the clusters in the value parameter space. Since cluster $0$ is made up of two distinct sets of points, we performed two simulations for cluster $0$ near the centroids of these sets. We call the set in between clusters $1$ and $2$ $0v$ and the other set $0s$.  Fig. \ref{vissims} visualizes the results: Set $0v$ of cluster $0$ represents a transitional state between chaotic and wave-like spreading types where some regularity remains, but the wave fronts have become very irregular. We call this type of behaviour the broken wave. Set $0s$ represents behaviour where the wave fronts strongly break in the diagonal directions, but are almost completely wave like in the cardinal directions. Cluster $2$ represents wave-like spreading pattern, while cluster $1$ represents the fully chaotic spreading pattern, as expected. It is interesting that the diagonally broken wave of set $0s$ seems by the naked eye to have more in common with the wave cluster $2$, but still classified by SOM as part of the broken wave cluster $0$. However, as we noted above, there are reasons to believe that the transitional cluster $0$ contains many different subdivisions with different behavioural types, and it would take a lot more detailed study to identify them all. 

The time-evolution of the district infection rates of these simulations, averages of which the present classification is based on, are depicted in Fig. \ref{infrates}. The wave-like cluster $2$ exhibits naturally very regular periodic behaviour, while the chaotic cluster $1$ has not even fully settled into a regular periodic pattern. Cluster $0s$ also has very regular behaviour like cluster $2$, but with noticeable differences in the evolution of the infection rates in the central district $5$ and cardinal districts $4$, $6$ and $8$: Instead of having a single seasonal peak infection rate, the primary peak in these districts is followed by a secondary peak shortly after the primary. Also, the peak infection rate in the central district is significantly smaller than in cluster $2$ and in general the evolution of the infection rates are not as smooth in the districts $4$, $6$ and $8$ as the infection waves subside. Interestingly, though, the behaviour is very similar in the diagonal districts $1$, $3$, $7$, and $9$ in clusters $2$ and $0s$. the evolution of the infection rates in cluster $0v$ seem very similar to those in cluster $1$, except much more rapid and coherent across districts with similar geographic positions relative to the origin of the epidemic.

The reason for the transition from wavelike spreading patterns to chaotic ones at $\Tilde{w}^c = 1$ can be found in Eq. \ref{dmeqgovmk}, where $(\Tilde{w}^c - 1)^{-1}$ appears as a coefficient. In the simulations the value parameters are randomized for each population agent so that $0.9 \le \Tilde{w}^c \le 1.1$ in the vicinity of the $\Tilde{w}^c = 1$ point. this means that $(\Tilde{w}^c - 1)^{-1}$ can have both positive and negative values, and that those values have very large absolute values. In fact, \begin{equation}
\lvert (\Tilde{w}^c - 1)^{-1} \lvert \ge 10
\end{equation}
always. The end result is that in this limiting case all the population agents are either doing their utmost or very little, if anything at all, to mitigate the spread of the epidemic, and it is random which behavioural type any given agent adopts. In the simulations this manifests as the characteristic broken wave shown in the upper right panel of Fig. \ref{vissims}.

The reason why cluster $0s$ mainly consists of simulations with small $\lvert \Tilde{W}^Y \lvert$ and large $\lvert \Tilde{w}^y \lvert$ is most likely a combination of following factors: 
\begin{enumerate}

\item  $\Tilde{w}^c$ is negative for all simulations in cluster $0s$. This means that population agents will respond negatively to pandemic regulations.

\item  $\Tilde{W}^Z$ is positive for all simulations in cluster $0s$. Positive $\Tilde{W}^Z$ means above average regulations (as we will see below), disincentivising the action by population agents in the first place.

\item  Large $\lvert \Tilde{W}^Y \lvert$ also mean tougher regulations, and so cluster $0s$ will be limited to small values of $\lvert \Tilde{W}^Y \lvert$.

\item $\Tilde{w}^y$ on its own encourages the population agents to act for epidemic mitigation, and so $0s$ will consist of simulations with large values of $\lvert \Tilde{w}^y \lvert$.

\end{enumerate}

In other words, the most likely explanation for the placement of the cluster $0s$ in the value parameter space is that the simulations are placed in this class by SOM if in these simulations the health consciousness of the population agents is strong enough to cause significant enough disruptions to the wavelike spreading patterns of the epidemic despite the aversion the agents feel towards complying with the epidemic regulations. 

Considering that the idea that governments would like to have their economy harmed by their own regulations, which is implied by $\Tilde{W}^Z > 0$, we also made SOM classifications using only the simulations with $\Tilde{W}^Z < 0$ as an experiment. The results of this experiment are shown in an appendix, and for the infection rate and $\beta$ feature vectors they are striking in that in these halved parameter spaces there are three clear clusters to be found that clearly correspond to the three main behavioural types of the model detailed above. In essence, in this limited space the complications brought by subcluster $0s$ disappear. 

\begin{figure}[t]
\centering
\epsfxsize = 0.45\columnwidth \epsffile{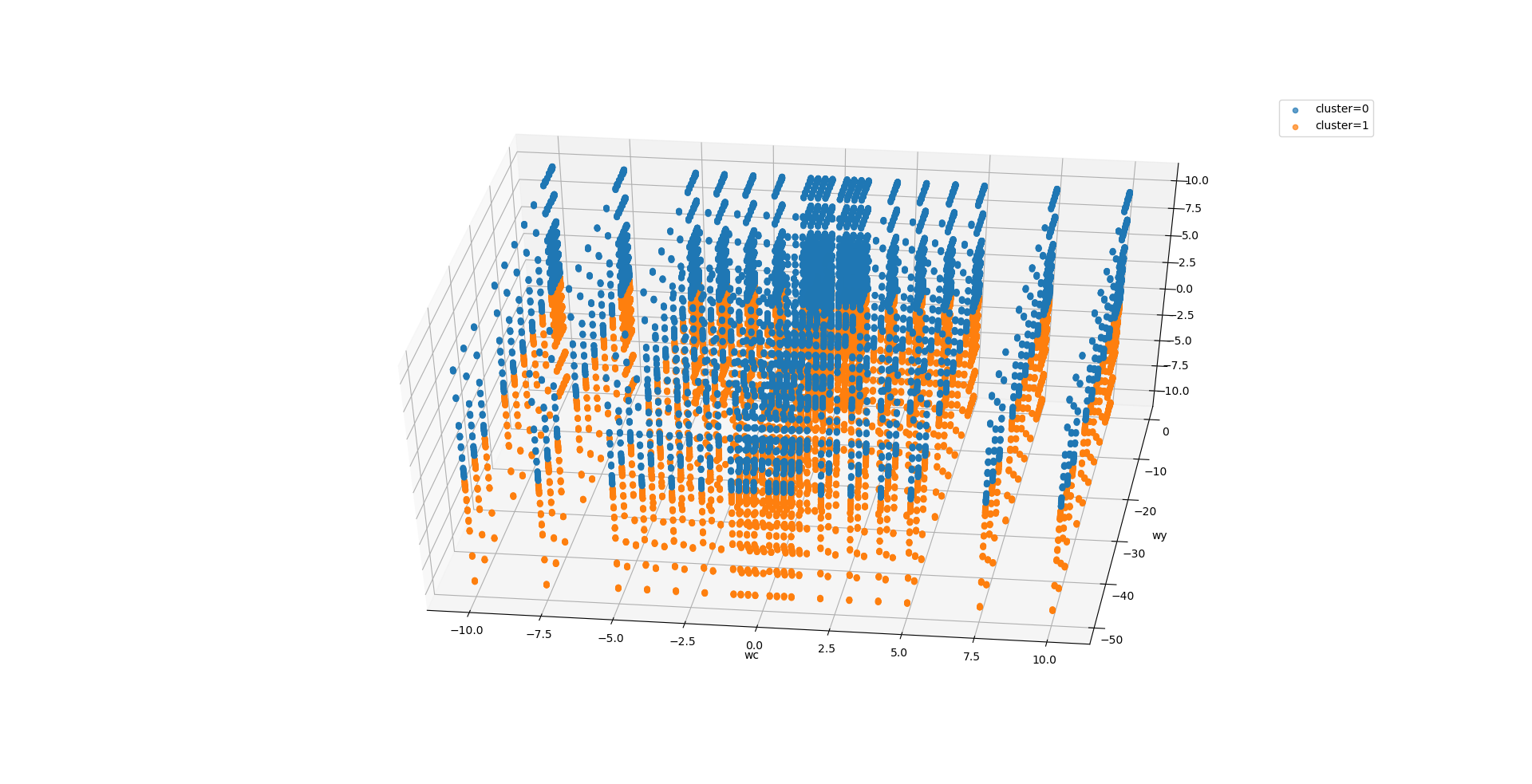}
\epsfxsize = 0.45\columnwidth \epsffile{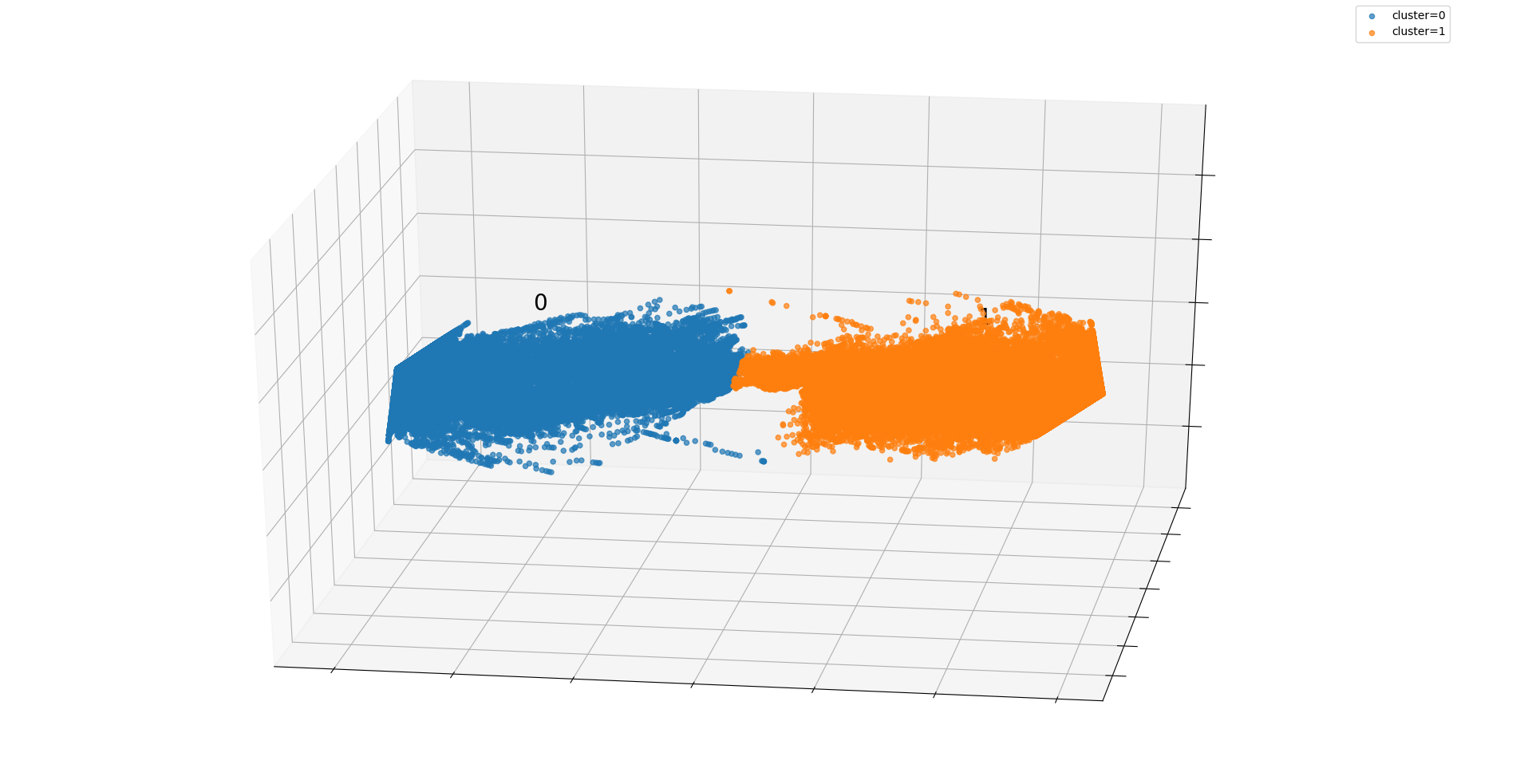}
\epsfxsize = 0.45\columnwidth \epsffile{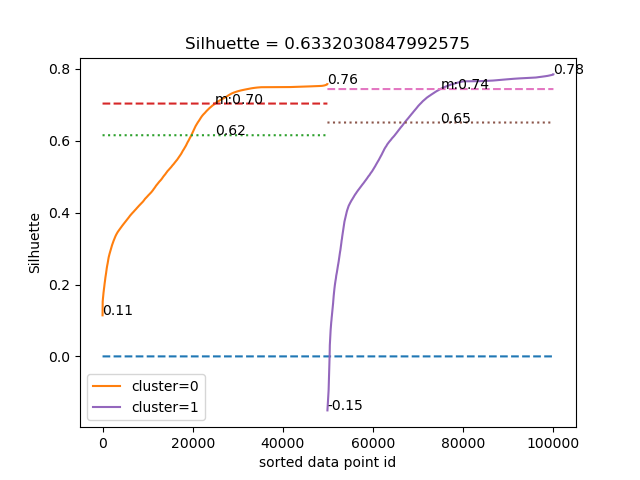}
\epsfxsize = 0.45\columnwidth \epsffile{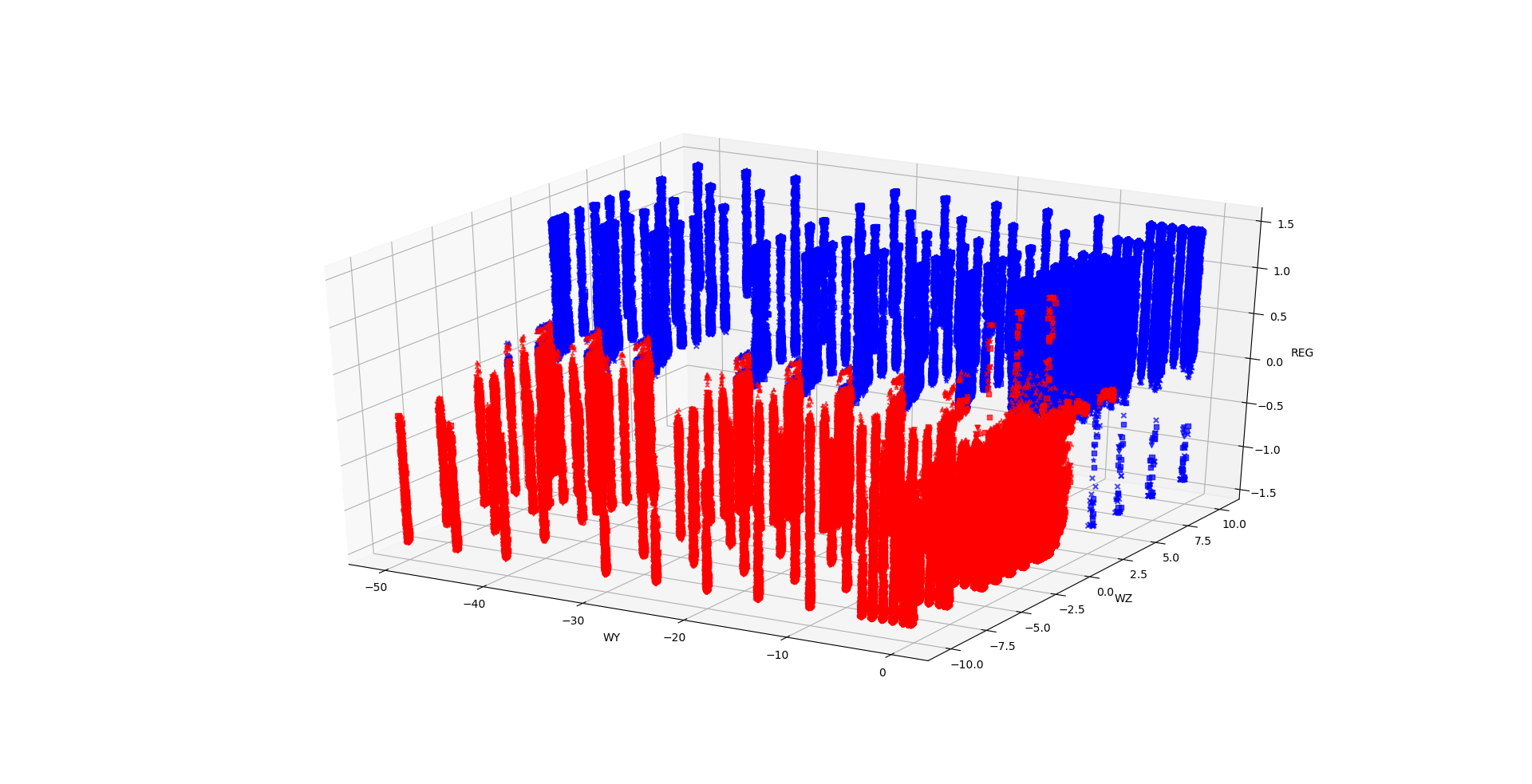}
\caption{The classification using a two neuron SOM with the average government regulations as the feature vector. Displayed are the classification in the value parameter space (top left panel) and in PCA space (top right panel), along with the silhouette profile (bottom left panel) and a three dimensional scatterplot of the simulated districts with normalized regulations on the vertical axis and value parameters $\Tilde{W}^Z$ and $\Tilde{W}^Y$ on the horizontal axes (bottom right panel). The axes in the value parameter space are the same as in lowest left panel of Fig. \ref{Sompcafig1}. }
\label{regfig}
\end{figure}

\subsection*{Regulation classification}

Judging from Figs. \ref{pcafig1} and \ref{silfig1}, the  governmental regulation classification also gave reasonable results, at least with only two SOM neurons. In Fig. \ref{regfig} we present the results of this classification: the identified clusters in the value parameter space (top left panel) and in PCA space (top right panel), the silhouette profile (bottom left panel) and a three dimensional scatterplot of the regulations in the simulated districts with normalized regulations on the vertical axis and value parameters $\Tilde{W}^Z$ and $\Tilde{W}^Y$ on the horizontal axes (bottom right panel). It turns out that in the value parameter space the two clusters identified by SOM occupy the upper (cluster $1$) and lower halves (cluster $0$) of the $W^Z = 0$ plane, symmetrically, while in the PCA space we have the expected result in which each of the visible clusters have been seemingly accurately identified by SOM. While the maximum silhouette numbers are very large for both of the clusters, around $0.77$, there are also relatively large numbers of simulations with relatively low silhouette numbers, especially in cluster $0$, while only cluster $1$ has some negative silhouette numbers. Thus, even at its best, the regulation classification does not quite have the quality of the infection rate and $\beta$ classifications presented above, which had clusters with much steeper dips in silhouette numbers and longer plateaus.

The scatterplot of the regulations vs. the economic and infection rate value parameters tells us that the main difference between the regulatory behaviours of governments in these clusters is that in the one with positive $\tilde{W}^Z$ the regulations tend to be higher than average, while in the other cluster the regulations generally fall below the average, with some overlap probably due to a presence of a bridge between the clusters visible in the PCA space. The $\tilde{W}^Y$ parameter turns out to have much less effect on the regulations, except when its values are near $0$, where some districts in some simulations have either above or below the average regulations relative to the clusters they belong. The differentiation of the simulations into two classes according to whether they have positive or negative $\tilde{W}^Z$ is to be expected, since in Eq. \ref{dmeqgovmkII} the term associated with $\tilde{W}^Z$ governs the competition of the authority agents between each other in economic hit their districts are taking: positive $\tilde{W}^Z$ means that the authority agents view the declining economy positively, and even compete with others to drive it down further. Whether or not this sort of outlook is actually shared by any of the existing real world authorities is questionable, of course, but in this study we have included this range of value parameters for the sake of completeness. If we consider the regulation classification only for the simulations with $\Tilde{W}^Z < 0$, the left hand cluster seen in the upper right panel of Fig. \ref{regfig} simply disappears, and we are left with only one recognisable cluster  (see appendix).

\section*{Discussion}

In this study we have performed a survey of the value parameter space of the hybrid socioeconomic epidemic BTH--SEIRS model we presented in \cite{SBKK2022}. We have done this by wrapping two of the value parameters, $W^X$ and $w^x$, into others and varying the resulting value parameters $\Tilde{W^Z_i}$,  $\Tilde{w^c_i}$, $\Tilde{W^Y_i}$ and  $\Tilde{w^y_i}$ for all the agents in the vicinities of $50 000$ grid points in the value parameter space with two simulations each, or $10^5$ simulations in total. We have then used a variety of computational classification tools to analyse this vast data: SOM to classify the data, PCA to help visualize the feature vectors used in the SOM classifications, and silhouette numbers to evaluate the quality of the SOM classifications. The feature vectors of our classifications are based on the averages of the four main quantities tracked by the model, which are infection rates, economic activity, government regulations and the popular compliance, and a special $\beta$ factor we introduced in order to measure the propagation speed of the epidemic. The main motivations of this study is to explore different phases in the behaviour of the model, and transitions between those phases, and to develop computational tools for the future applications of the model to real world situations. 

Our main result is that the SOM classifications based on average infection rates produce results that are most easily interpreted physically. SOMs recognize the main behavioural types of the model that we already identified in \cite{SBKK2022}, the chaotic and wave spreading patterns, and additionally find the transitional areas in the parameter space between these two behavioural types. SOMs also reveal that the transition from one type to another is very complex in the value parameter space, as judged by the fact that increasing the number of neurons when running the SOM results in the transitional cluster being subdivided into more and more subclusters. The most significant value parameter in terms of the epidemic spreading type turned out to be the one representing the compliance of the general populations $\Tilde{w^c_i}$, with $\Tilde{w^c_i} = 1$ marking the transitional state between the wavelike spreading patterns with $\Tilde{w^c_i} < 1$ and the chaotic spreading patterns with $\Tilde{w^c_i} > 1$. The classifications based on the $\beta$ factor agreed with the infection rate classifications very closely.

The classifications based on government regulations were second most easily interpretable, although they were even simpler than the infection rate classifications, as we only detected only two clusters with PCA in this case. The differences between these clusters were straightforward: one cluster had  negative $\Tilde{W^Z_i}$ (government concern for the economy of their district) and below average government regulations, while the other had positive $\Tilde{W^Z_i}$ and above average government regulations. Other value parameters did not have significant effects on this classification. What makes this case somewhat less easy to understand than the infection rate and $\beta$ classifications is the fact that positive $\Tilde{W^Z_i}$ implies that the government sees the decline of the economy as a positive development, which is not how most real modern governments view things, with the possible exceptions of China and North Korea. We note, however, that in the long sweep of history different societies and their governments have had very different views to economic growth than the modern society, and that the economic attitudes can easily change again in the future. Therefore, we have chosen to study the  $\Tilde{W^Z_i} > 0$ case for the sake of completeness. Otherwise, result obtained from the the regulation classification is simple: if the governments are not afraid of causing harm to their economy, they will use tougher measures to mitigate the epidemic.

While we attempted to use the same classification methods to economic activity and popular compliance, the PCA results did not indicate the presence of clear cut clusters present in these feature vectors, and so we did not analyse them further. More specifically, PCA showed that these feature vectors consisted of a bunch of one dimensional strings forming singular thick bundles. While it may be possible to find more detailed structure in these cases using more advanced data analysis techniques such as UMAP or t-SNE, we chose to limit ourselves to simple tools such as PCA and SOM, because the BTH--SEIRS model we were testing is still in proof of concept stage, and to dive too deep into its present form might be wasted effort.

The future work on for BTH--SEIRS model will be aimed at applying it in a real world situation. This study has provided new insights into the behaviour of the model that will be useful in the effort, as well as provided a testing ground for the computational tools that may be necessary for such applications. While the simple form of the model we used in this study could be analysed further, for example by allowing greater regional variation than we have, we see little use for excessive testing of the proof of concept at this stage.

\bibliography{BTHgeneral}
\newpage
\appendix{ Appendix PCA and silhuette results}

\begin{figure}[h!t]
\centering
\epsfxsize = 0.45\columnwidth \epsffile{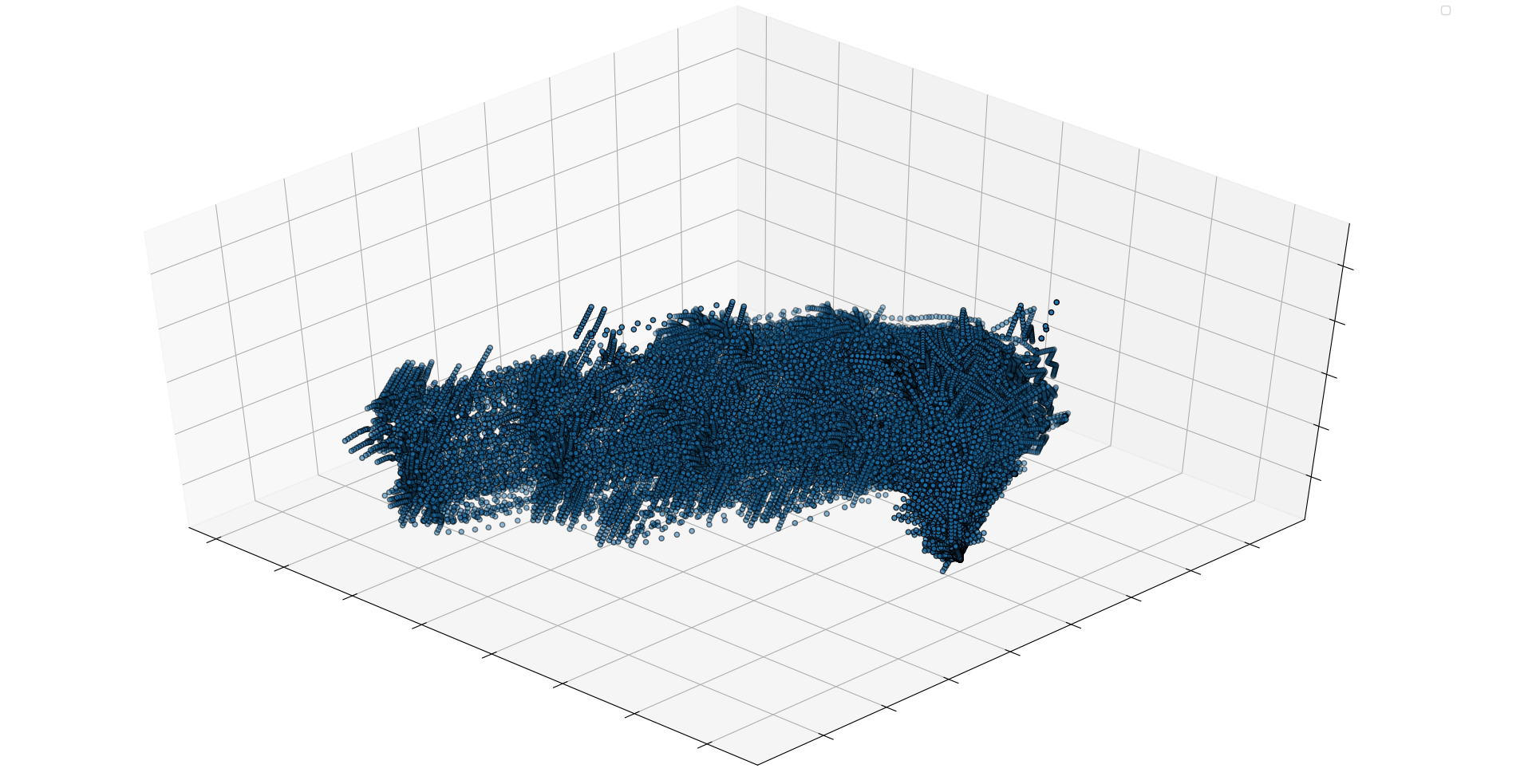}
\epsfxsize = 0.45\columnwidth \epsffile{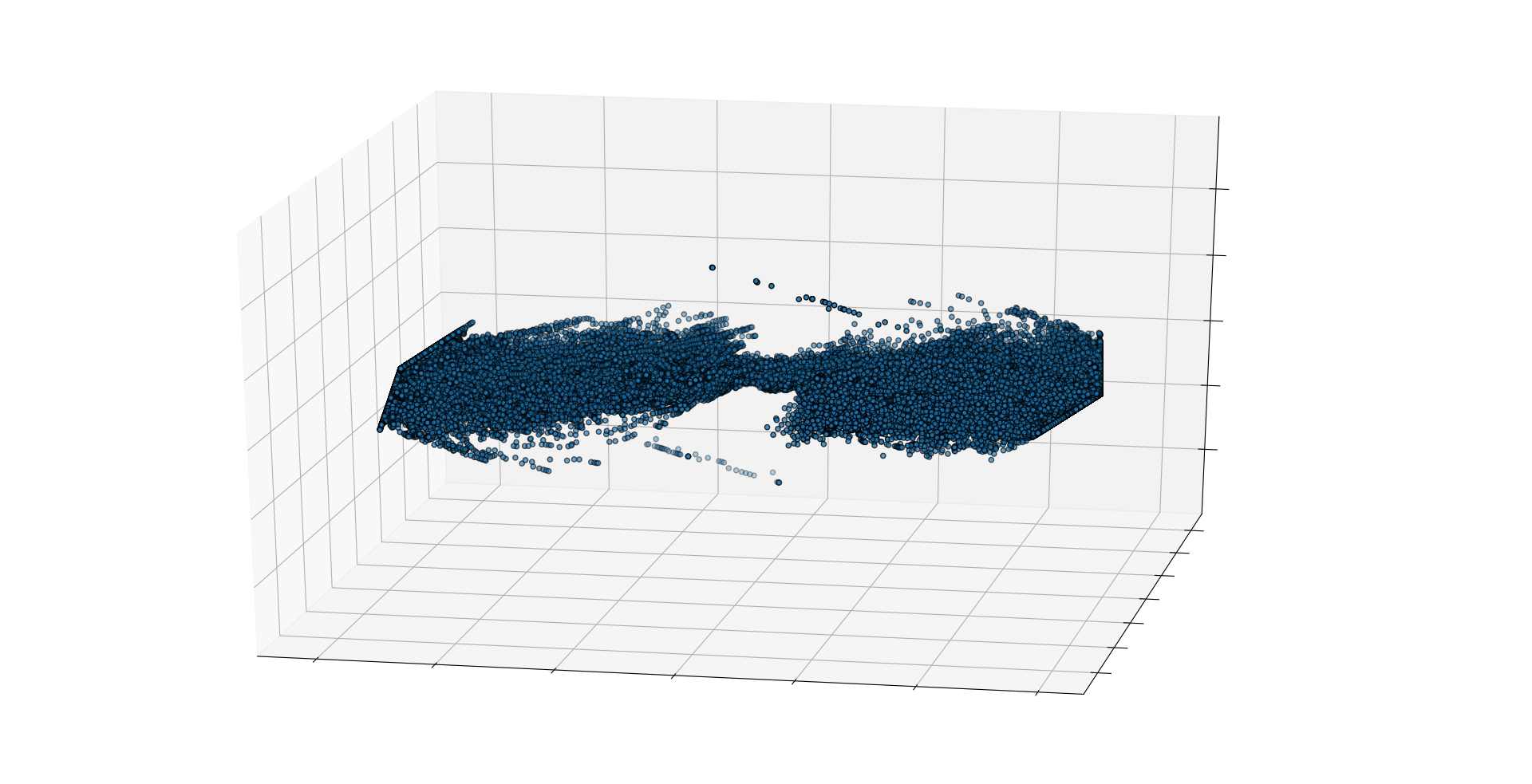}
\epsfxsize = 0.45\columnwidth \epsffile{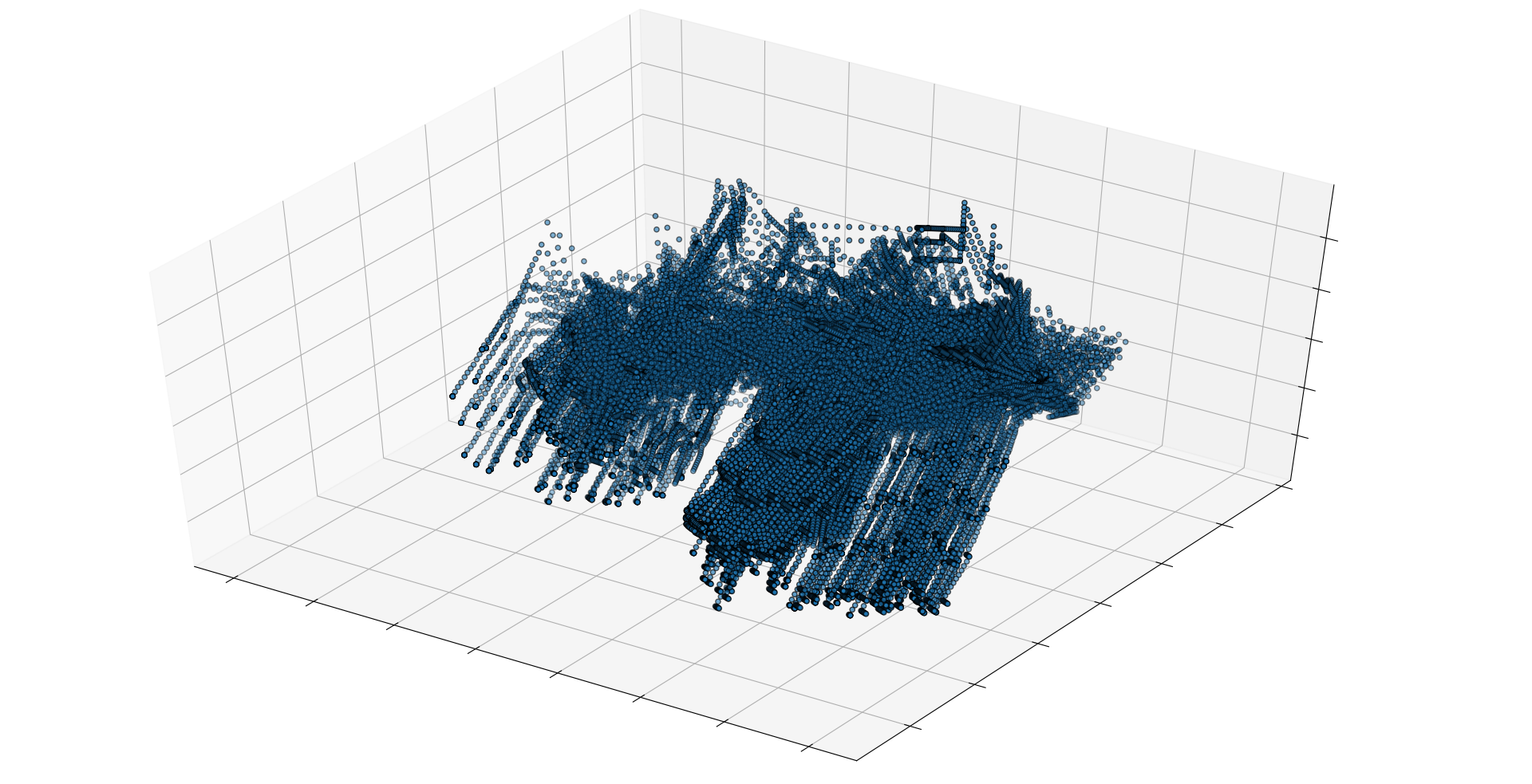}
\epsfxsize = 0.45\columnwidth \epsffile{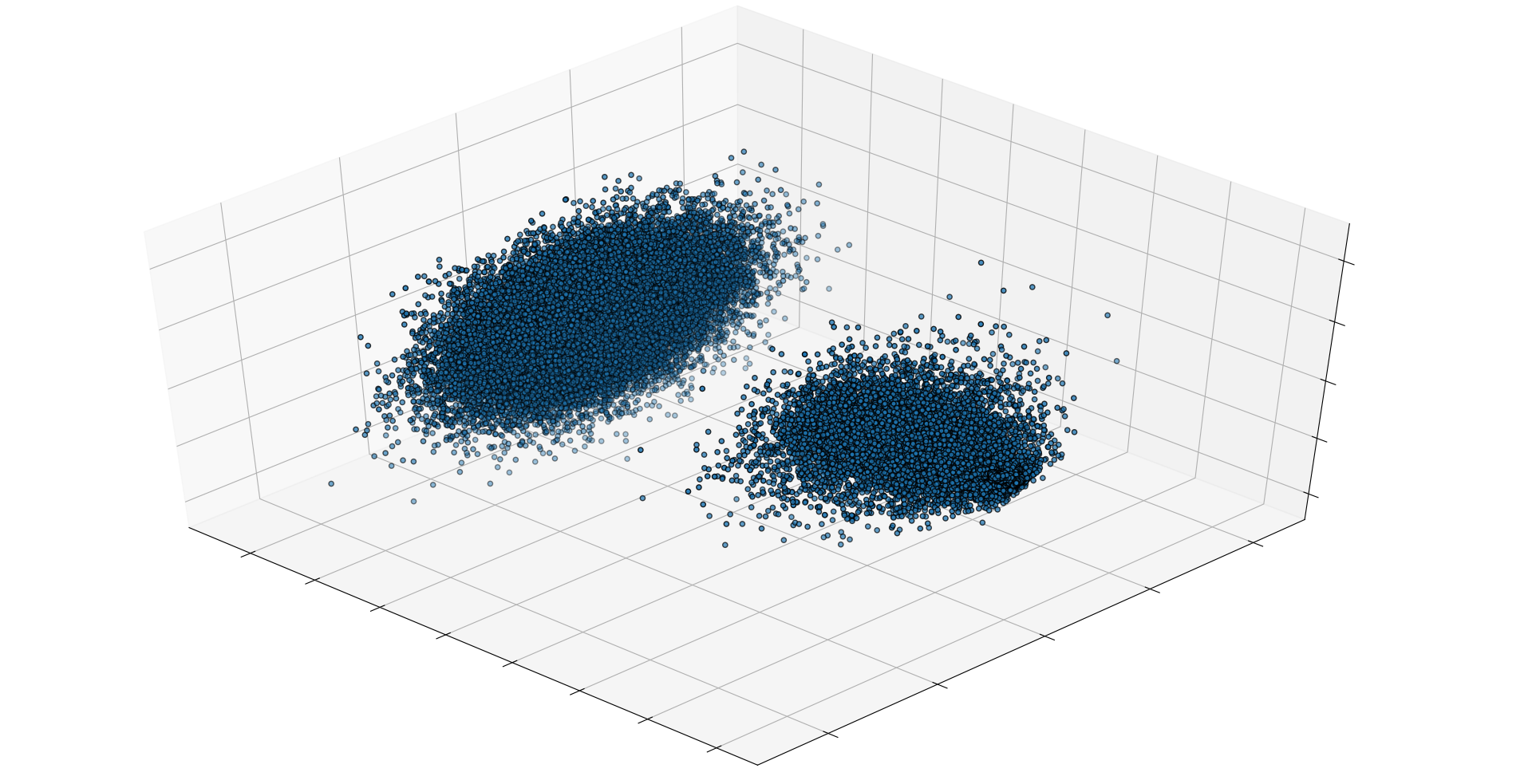}
\epsfxsize = 0.45\columnwidth \epsffile{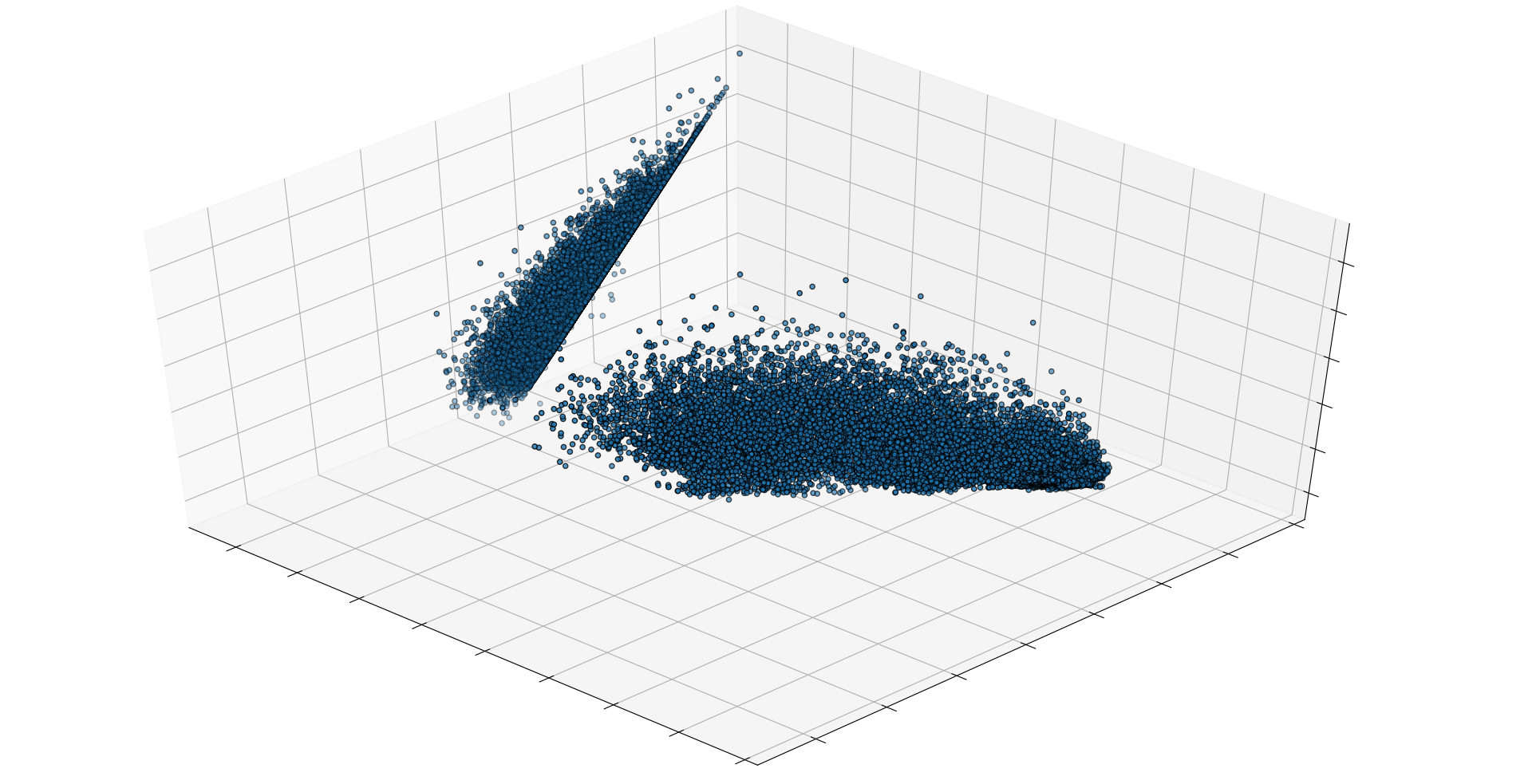}
\caption{The PCA profiles of the feature vectors constructed from the average economic activity (upper left panel), government regulations (upper right panel), dynamic compliance (middle left panel) and infection rates (middle right panel) and the $\beta$ measure (lowest panel). }
\label{pcafig1}
\end{figure}

 Fig. \ref{pcafig1} shows the results of the PCA analysis of our data with these feature vectors. It turns out that the economic feature vector produces a shape in the PCA space that does not readily lend itself to classification, at least when judged with human eyes, as there is only one easily identifiable cluster with rather complicated structure and and with some filamentary features. This is also the case for the compliance feature vector, although there are a lot more filamentary features. The feature vector made from the average governmental regulations looks like two clusters of points touching each other, with interesting symmetrical filaments attached to each cluster. The feature vector constructed from the average infection rates seem the easiest to classify, as it takes the form of two easily distinguishable clusters, one more dense than the other. The $\beta$ measure feature vector, similarly, consist of two easily identifiable clusters, but the shapes of the clusters are very different from the infection rate feature vector: they are almost winglike in shape, one more dense than the other.

\begin{figure}[h!t]
\centering
\epsfxsize = 0.45\columnwidth \epsffile{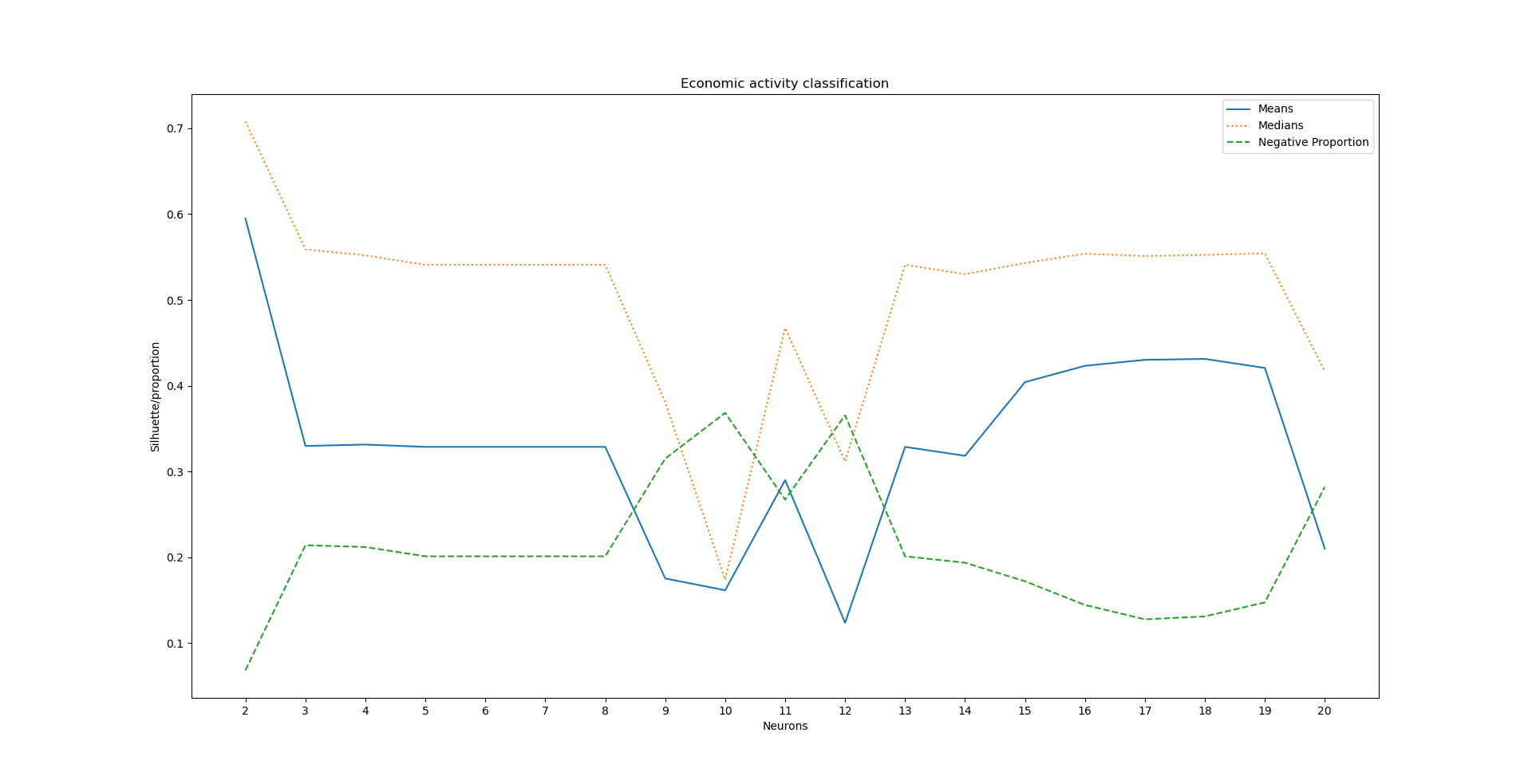}
\epsfxsize = 0.45\columnwidth \epsffile{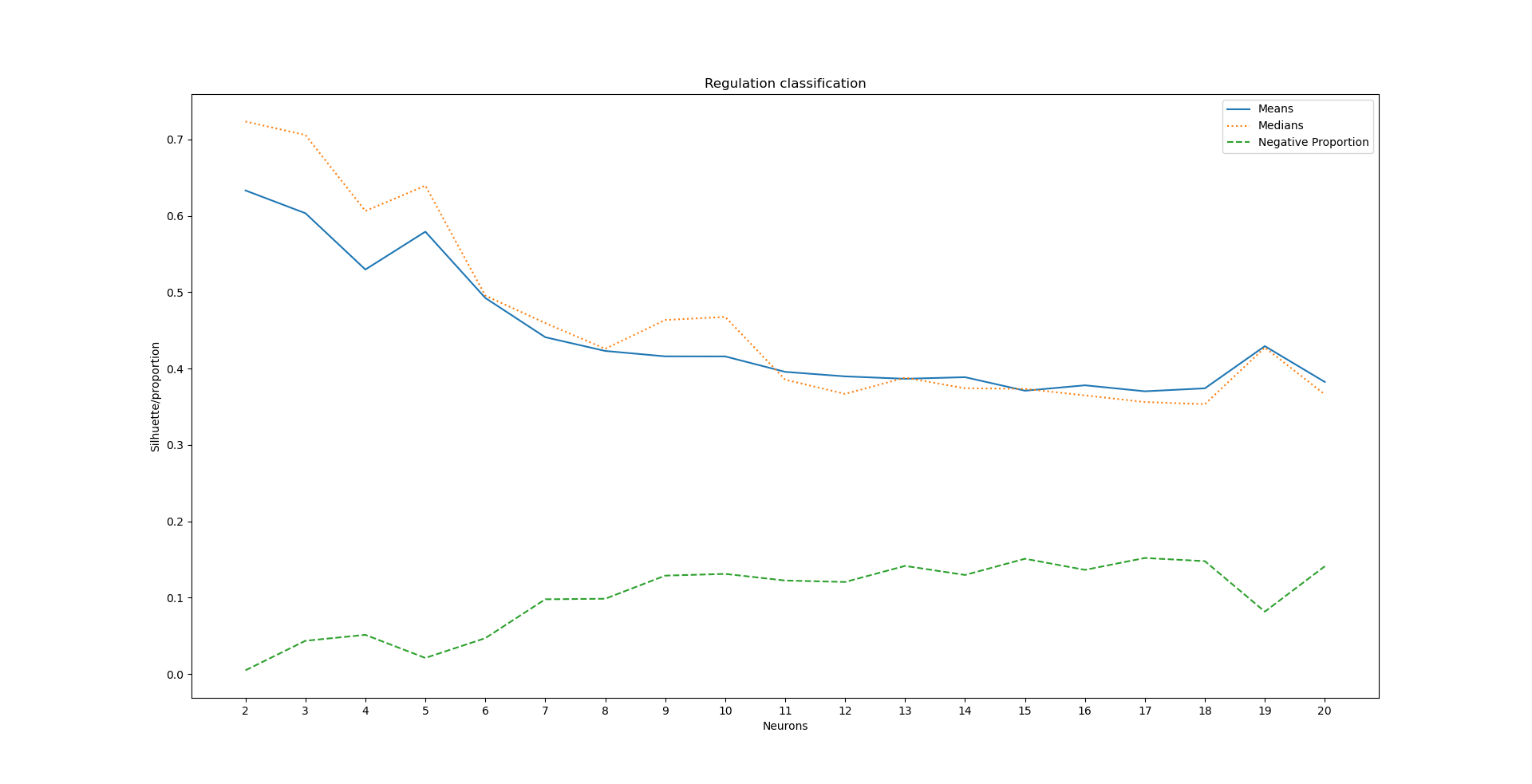}
\epsfxsize = 0.45\columnwidth \epsffile{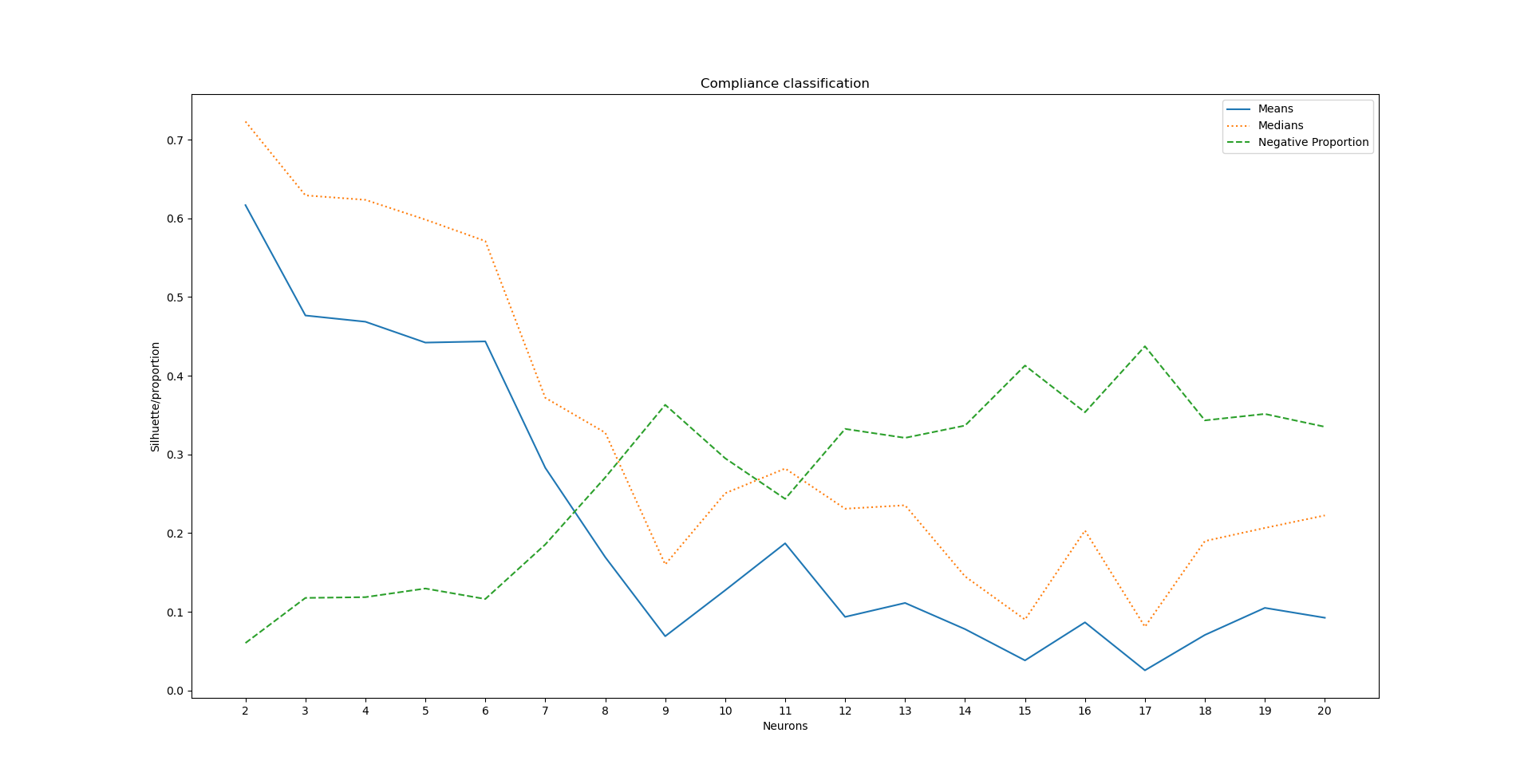}
\epsfxsize = 0.45\columnwidth \epsffile{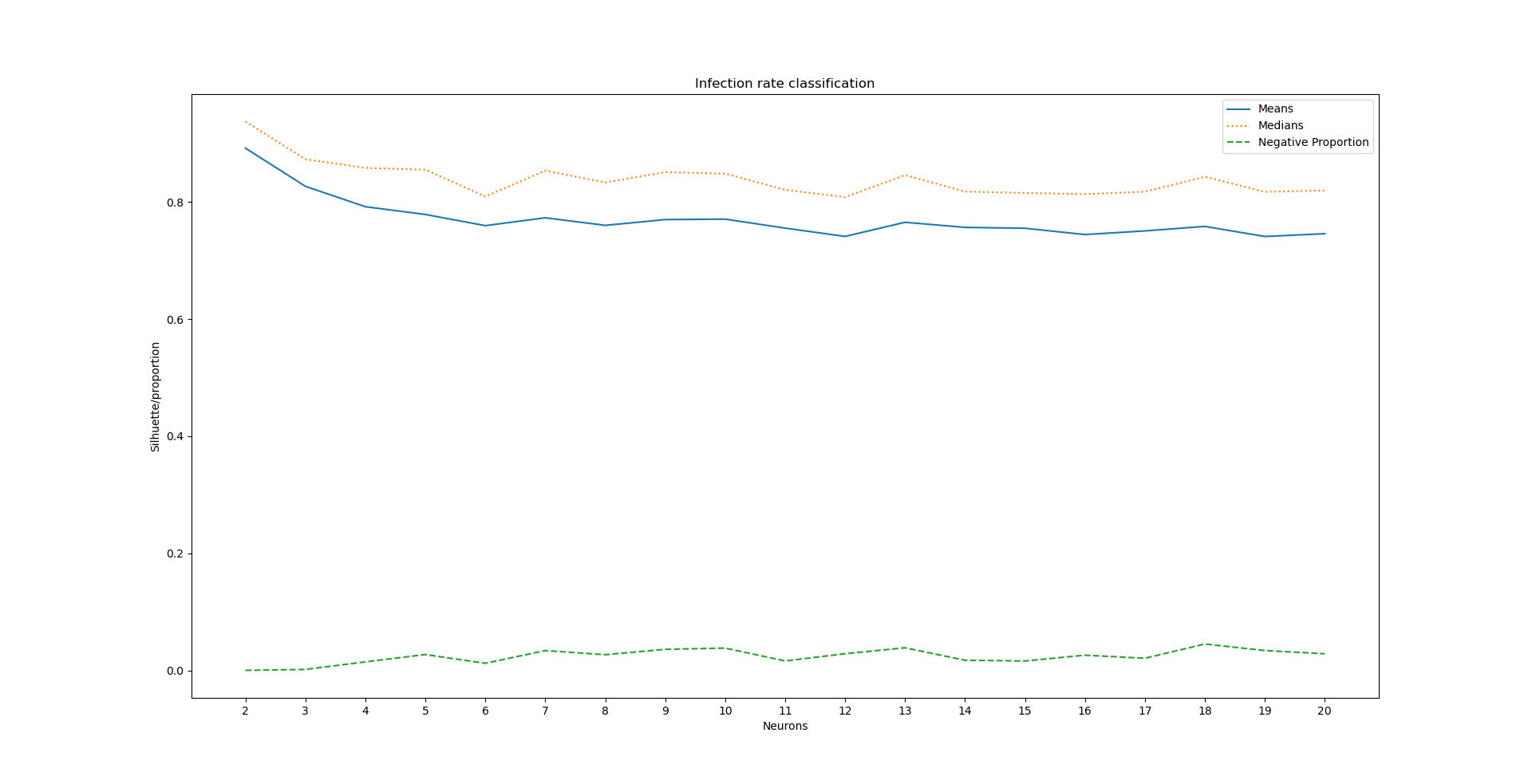}
\epsfxsize = 0.45\columnwidth \epsffile{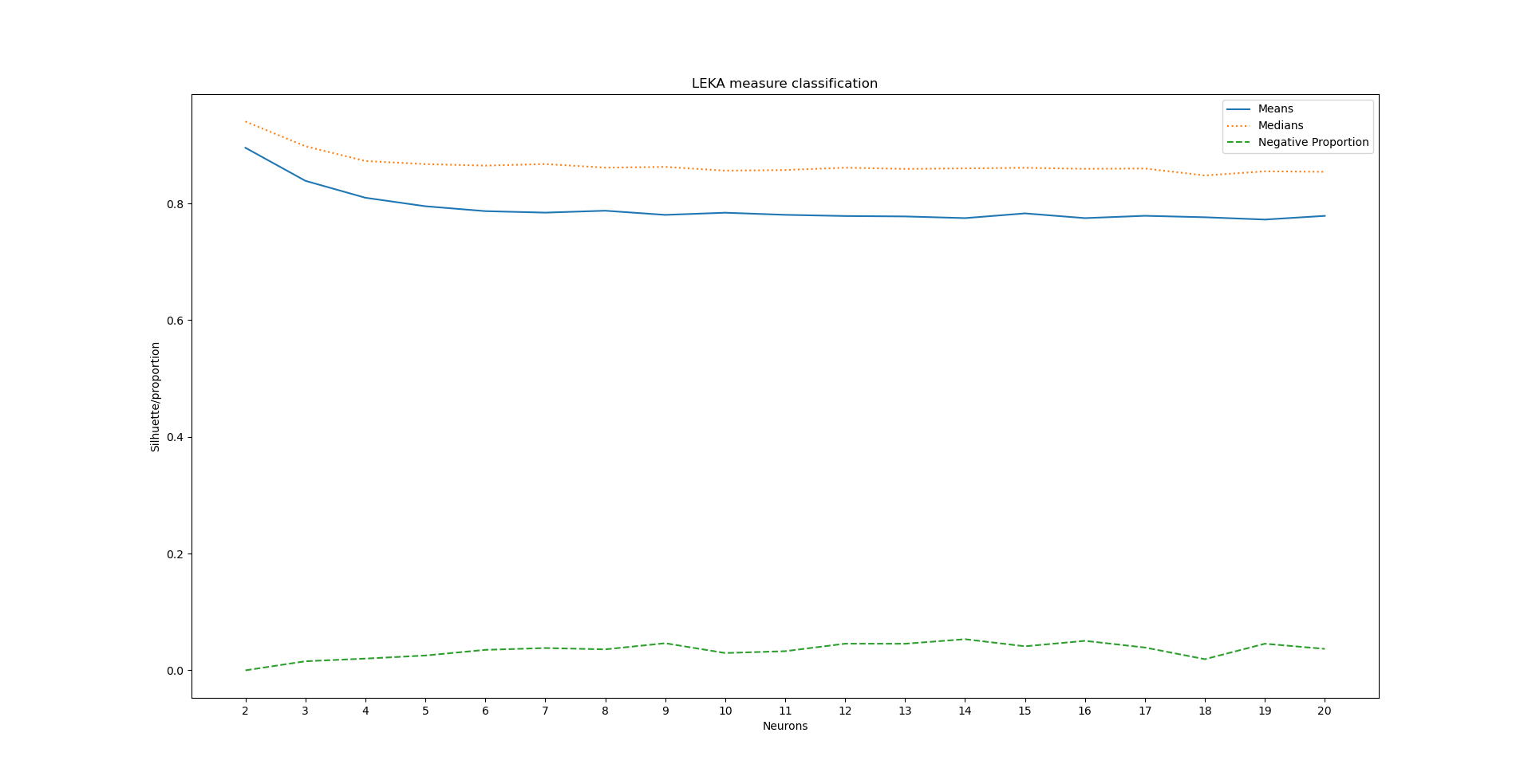}
\caption{The mean and median silhouette numbers, along with the proportion of negative silhouette numbers when SOMs are used to classify the simulations according to the feature vectors visualised in Fig. \ref{pcafig1}. }
\label{silfig1}
\end{figure}

It is clear from the PCA analysis that the feature vectors formed from the average governmental regulations, the infection rates and the $\beta$ measures are most amenable for computational classification. In this study we use SOM classification with $2$--$20$ neurons, and evaluate the quality of these classifications with standard silhouette numbers. Fig. \ref{silfig1} shows the average and median silhouette numbers with different SOM neuron counts and feature vectors, along with the proportion of the simulations with negative silhouette numbers in each classification. As expected, the silhouette numbers of the  classifications based on economic activity and dynamic compliance are much lower than those based on government regulations, infection rates and the $\beta$ measures. The compliance classification in particular has a clear downward trend in both median and average silhouette numbers, which fall from little over $0.7$ and  $0.6$ to about $0.2$ and $0.1$, respectively, while the proportion of negative silhouette numbers climbs from approximately $0.07$ to over $0.3$. 

The silhouette measures associated with the economic classification, in contrast, are seemingly fragmented according to the SOM neuron count into multiple different states: With two neurons, the median and average silhouette numbers are relatively large at about $0.7$ and $0.6$, respectively, which are proportional to those of the compliance classification at their highest. Similarly, the proportion of negative silhouette numbers are about the same. Between $3$ and $8$ neurons, the median and average silhouette numbers have relatively stable values at about $0.55$ and $0.33$, respectively, along with the proportion of negative silhouette numbers, which hover at little over $0.2$. Between $9$ and $13$ neurons all these measures rise and fall in zigzag pattern, until they end up at approximately their previous levels, except for the average silhouette numbers which zigzags until neuron count $15$ and reaches values higher than $0.4$, which is significantly more than before. Between neuron counts $13$ and $19$ the median silhouette numbers remain relatively constant, while the proportion of negative silhouette numbers have a slight decreasing trend, reaching a minimum well below $0.2$ at neuron count $17$. At neuron count $20$ both the median and average silhouette numbers fall significantly, and the proportion of negative silhouette numbers rises fast.

Of all the classifications, the ones based on the infection rates and the$\beta$ measure turn out to be by far the most well behaved. While there is a clear decreasing trend in both the average and median silhouette numbers, and a corresponding rising trend in the proportion of the negative silhouette numbers, the latter never climbs to more than approximately $0.12$, and the former never dip much below $0.8$. These trends are also present in the classifications based on the government regulations, but in much stronger form. While the proportions of the negative silhouette numbers are proportional to each other in these two classification schemes, the average and median silhouette numbers are in general noticeably smaller in the regulation classification than the infection rate classification, having maximum values of approximately $0.64$ and $0.73$, respectively, and minimum values of little over $0.4$.

\subsection*{Halving the parameter space: $\Tilde{W}^Z < 0$}

\begin{figure}[t]
\centering
\epsfxsize = 0.45\columnwidth \epsffile{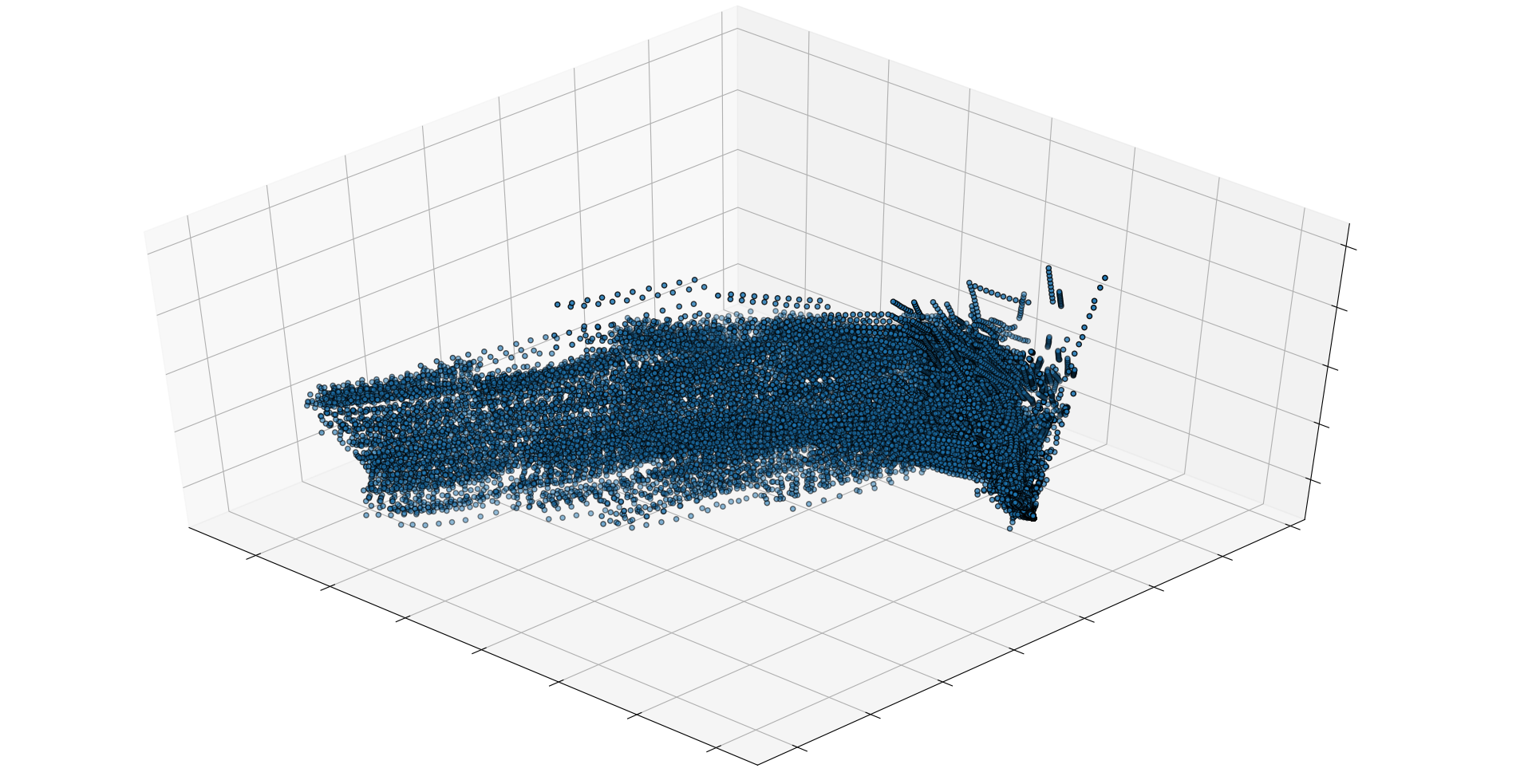}
\epsfxsize = 0.45\columnwidth \epsffile{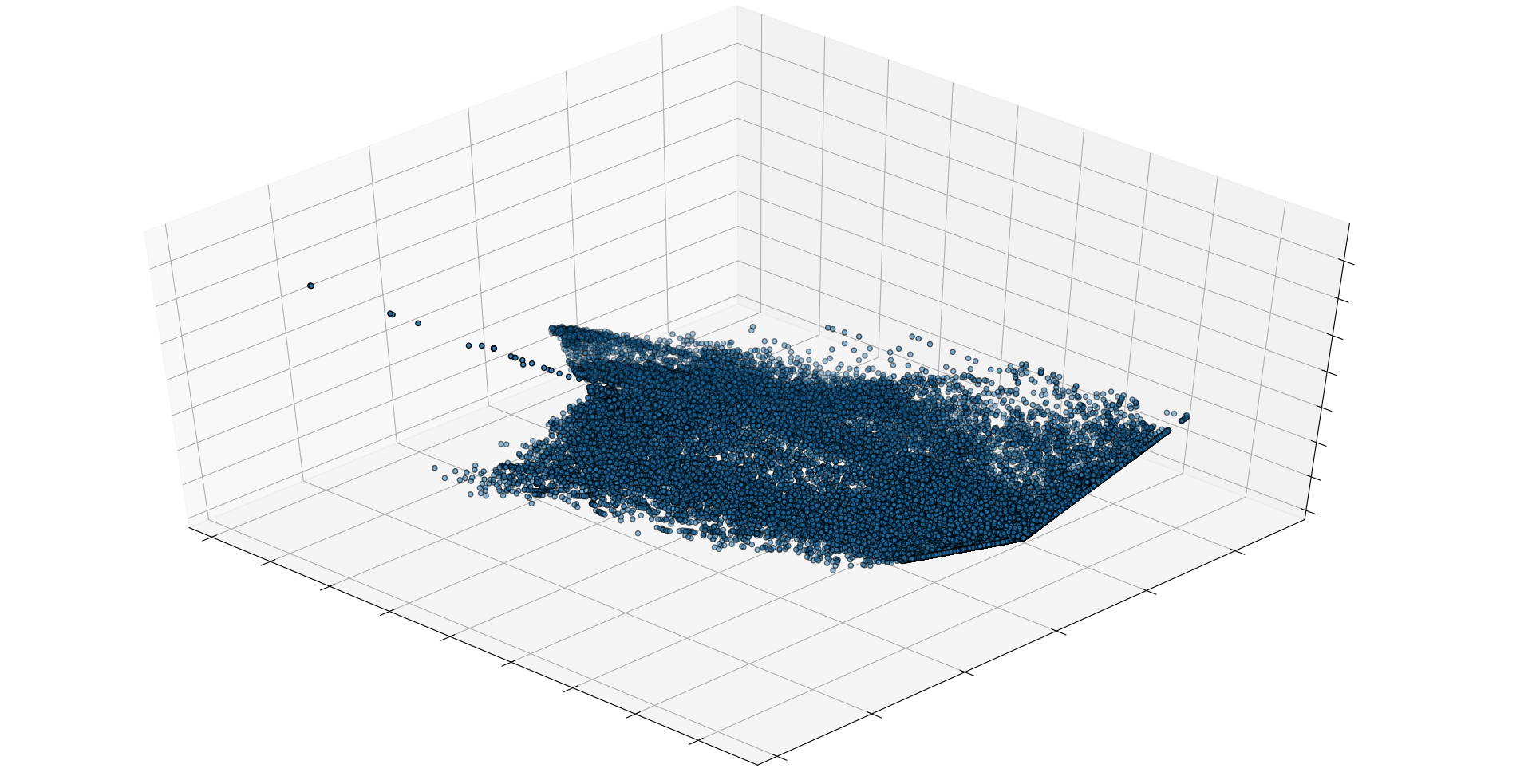}
\epsfxsize = 0.45\columnwidth \epsffile{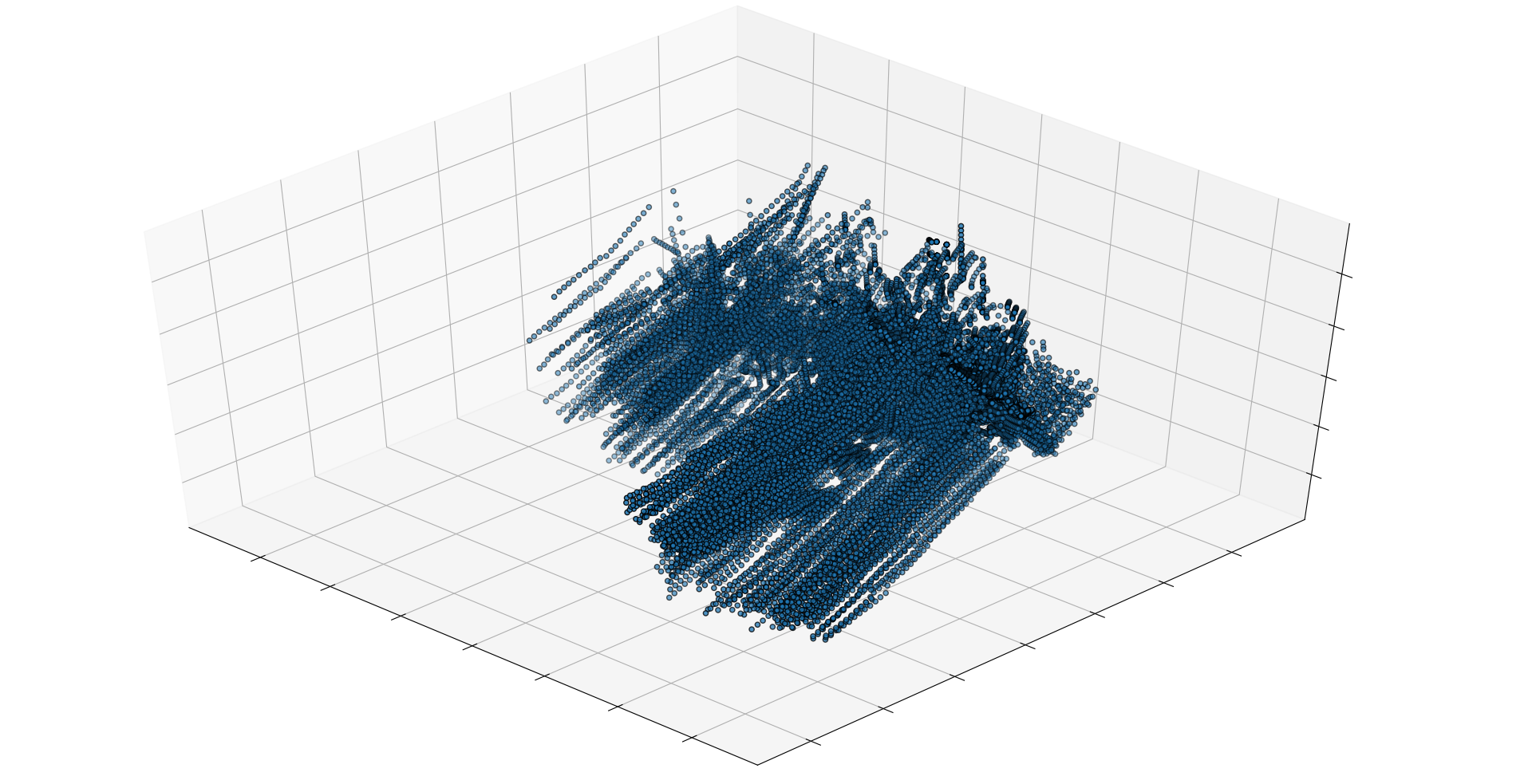}
\epsfxsize = 0.45\columnwidth \epsffile{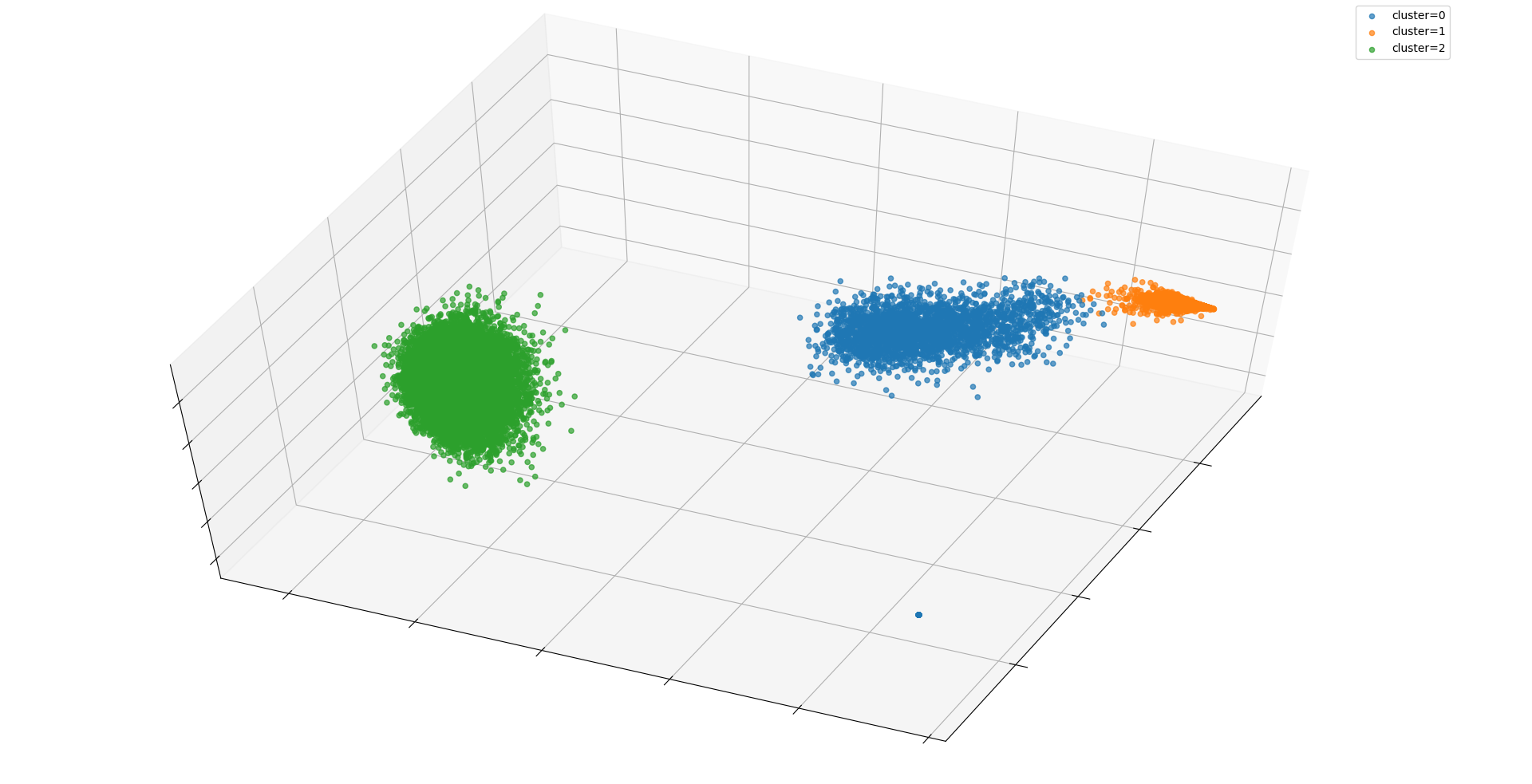}
\epsfxsize = 0.45\columnwidth \epsffile{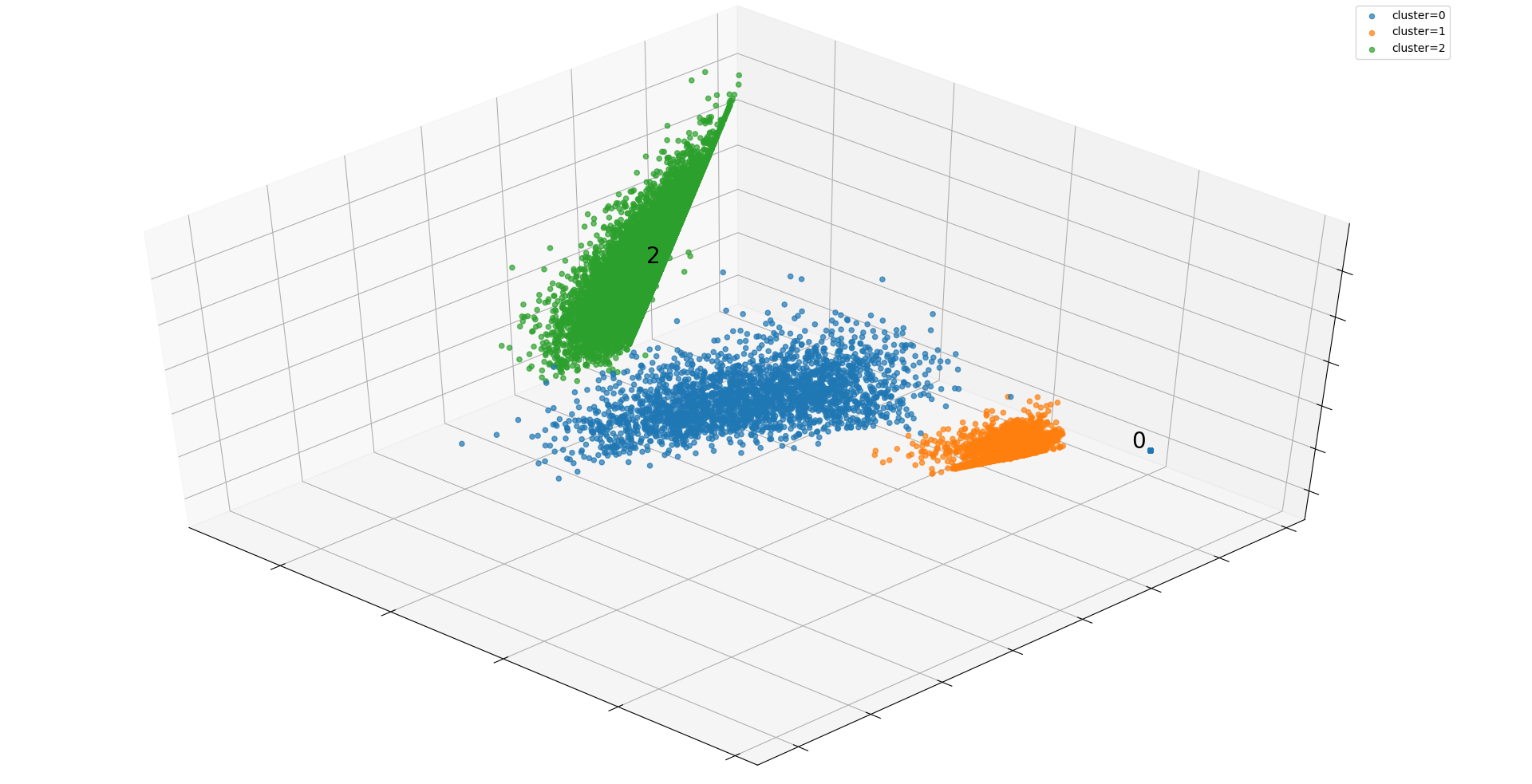}
\caption{The PCA profiles of the feature vectors when applied only to the half of the simulations with $\Tilde{W}^Z < 0$. The different feature vectors are presented in the same order as in Fig. \ref{pcafig1}. The $\beta$ and infection rate feature classifications have been classified using SOM into three clusters.}
\label{pcafigp}
\end{figure}

As noted in the main text, $\Tilde{W}^Z < 0$ means that governments are happy have their economies decline, which would be a very unusual attitude in the modern world. Because of this concern, we also applied our PCA and SOM procedures to the set of simulations with  $\Tilde{W}^Z < 0$ as an experiment. The results can be seen in Fig. \ref{pcafigp}, in which the feature vectors are represented in the same order as in Fig. \ref{pcafig1}, and with the infection rate and $\beta$ feature vectors classified by SOM into three clusters. The shapes of the economic and compliance classifications do not change much, while in the case of the regulation classification one of the clusters seen in Fig. \ref{pcafig1} disappears, which could be expected from Fig. \ref{regfig}. The most interesting difference, however, can be seen in the infection rate and $\beta$ classifications, in which three clearly distinct clusters emerge. These clusters, of course, represent the wavelike, broken wave, and chaotic behavioural patterns. Apparently, a part of the simulations with $\Tilde{W}^Z > 0$, most likely the subgroup $0s$, forms a bridge between the broken wave and fully wavelike patterns in the full PCA figure (Fig. \ref{pcafig1}), thereby obscuring the simpler clustering seen in \ref{pcafigp}. The clusters shown in the PCA space in \ref{pcafigp} position themselves in a predictable manner in the value parameter space: wavelike forms occur when $\Tilde{w}^c < 1$, broken waves when $\Tilde{w}^c \approx 1$, and chaotic forms when $\Tilde{w}^c > 0$.

\end{document}